\documentclass{article}
\usepackage{arxiv}
\usepackage{graphicx} 
\newif\ifclean
\cleanfalse 
\usepackage{graphicx}
\usepackage{cite}
\usepackage{amsmath}
\usepackage{amssymb}
\usepackage{subcaption}
\usepackage{algorithm}
\usepackage{algpseudocode}
\usepackage{amsthm}
\usepackage{framed}

\usepackage{hyperref}
\usepackage{listings}

\graphicspath{{Figures/}}
\usepackage{comment}

\usepackage{soul}
\usepackage{todonotes}
\ifclean
\newcommand{\COMMENT}[1]{{}}
\newcommand{\QUESTION}[1]{{}}
\newcommand{\TODO}[1]{{}}
\else
\usepackage[dvipsnames]{xcolor}
\usepackage{soul}
\newcommand{\COMMENT}[1]{\textcolor{cyan}{{[ \sc{#1} ]}}\\} 
\newcommand{\QUESTION}[1]{\textcolor{blue}{{[{QUESTION:} \it{#1} ]}}\\} 
\newcommand{\TODO}[1]{\textcolor{red}{TODO: {\bf{#1}}}\\} 

\newcommand{\red}[1]{\textcolor{red}{{#1}}}
\newcommand{\caution}{\red{\bf Draft: \today. Do not distribute.}}
\fi

\usepackage{upgreek}

\newcommand{\cref}[1]{Ref.\,\cite{#1}}

\usepackage[T1]{fontenc}

\renewcommand{\TODO}[1]{\todo{{#1}}}

\title{Full-Field Calibration of Coupled Thermomechanical Material Models at Finite Strain}
\renewcommand{\shorttitle}{\bf Full-field thermomechanical calibration}
\author{
}
\date{\today  }

\author{
L. River Spencer \\
Department of Aerospace Engineering \& Engineering Mechanics,\\
The University of Texas at Austin,\\
Austin, TX 78712 \\
\And
William D. Meador \\
Department of Biomedical Engineering,\\
The University of Texas at Austin,
Austin, TX 78712
\And
Adrian Buganza Tepole\\
Department of Mechanical Engineering, \\
 Columbia University,\\
New York, NY 10027\\
\And
Brian N. Granzow \\
Sandia National Laboratories \\
Albuquerque, NM 87185, USA
\And
Jin Yang\\
Department of Aerospace Engineering \& Engineering Mechanics,\\
Texas Materials Institute, \\
The University of Texas at Austin,\\
Austin, TX 78712 \\
\And
Manuel K. Rausch\\
Department of Aerospace Engineering \& Engineering Mechanics,\\
The Oden Institute of Computational Science and Engineering,\\
Department of Biomedical Engineering,\\
The University of Texas at Austin,\\
Austin, TX 78712 \\
\And 
D. Thomas Seidl \\
Sandia National Laboratories \\
Albuquerque, NM 87185, USA
\And
Jan N. Fuhg \\
Department of Aerospace Engineering \& Engineering Mechanics,\\
The Oden Institute of Computational Science and Engineering,\\
The University of Texas at Austin,\\
Austin, TX 78712 \\
}

\begin{document}
\setlength\parindent{0pt}
\maketitle

\footnotetext[1]{Corresponding author: \texttt{jan.fuhg@utexas.edu}}

\begin{abstract}
Calibrating thermomechanical material models from experiments is challenging because deformation, temperature, and force responses are strongly coupled, while measurements are usually restricted to specimen surfaces. We present a full-field calibration framework for coupled finite-strain thermomechanical material models using boundary displacement, reaction-force data, and temperature. The forward model is formulated as a near-incompressible thermo-hyperelastic problem with thermomechanical coupling derived from a Helmholtz free energy, and the inverse problem is posed as a PDE-constrained optimization problem with weighted observation terms for the available data streams.

Reduced gradients are computed with adjoint sensitivities that are obtained by automatic differentiation, enabling gradient-based calibration of nonlinear transient thermomechanical systems. The formulation is first verified on synthetic examples involving uniform thermal preconditioning and localized transient rod contact, where the ground-truth parameters are recovered from full-field measurements and force observations. The same workflow is then applied to experimental thermomechanical data by first calibrating a hyperelastic mechanical baseline from cyclic equibiaxial loading and subsequently identifying thermal expansion and directional shrinkage parameters from surface-temperature and boundary-force histories. The results demonstrate that coupled thermomechanical parameters can be inferred from experimentally accessible surface data without requiring volumetric observations.
\end{abstract}




\section{Introduction}
Thermomechanical effects directly govern the performance and reliability of soft materials such as polymers, elastomers, and biological tissues, which are widely used in applications ranging from biomedical devices to aerospace structures due to their lightweight and favorable mechanical properties. However, their response is strongly influenced by temperature, which affects stiffness, deformation, and long-term durability \cite{humphrey2003continuum,li2012molecular}. Thermal loading can induce significant changes in mechanical behavior, including transitions such as the glass transition in polymers and denaturation processes in biological tissues, where materials evolve from stiff and load-bearing to soft and compliant states \cite{chen1998phenomenological,wright2002denaturation}. In addition, soft materials typically exhibit relatively large coefficients of thermal expansion, making dimensional stability a critical design constraint, as temperature changes can lead to residual stresses and potential failure \cite{takenaka2012negative}. These effects are further amplified by thermomechanical coupling, where mismatches in thermal expansion or spatial temperature gradients generate internal stresses that strongly influence mechanical performance and failure mechanisms \cite{takenaka2012negative}.

From a physical perspective, the interaction between mechanics and temperature is governed by energy-conversion mechanisms, whereby mechanical work can be partially converted into heat or stored energy, thereby influencing the evolution of temperature and the material response. In biological tissues, for example, thermally activated processes such as collagen denaturation lead to shrinkage and changes in stiffness \cite{chen1997heat,chen1998heat}. As a result, many modern applications operate inherently under coupled thermo-mechanical conditions, necessitating models that capture both effects simultaneously \cite{humphrey2003continuum}. Moreover, thermomechanical behavior is highly sensitive to material composition and microstructure; in collagenous tissues, the underlying fiber architecture governs anisotropic responses such as directional shrinkage and force generation under thermal loading \cite{brodsky2005molecular,bancelin2015ex}.
Despite these coupled effects, experimental characterization of soft materials remains fundamentally limited. Classical calibration approaches based on homogeneous or uniaxial tests provide only integrated responses and fail to capture the spatially heterogeneous nature of thermomechanical behavior, particularly under multiaxial loading and finite deformations \cite{chen1998heat,harris2003altered}. Consequently, accurate predictive modeling requires inverse frameworks that account for the coupled interaction between deformation, temperature, and material response, rather than treating the mechanical and thermal problems in isolation \cite{humphrey2003continuum}.

To overcome these limitations, full-field measurement techniques have emerged as a powerful alternative. Digital image correlation (DIC) can provide spatially resolved kinematic information, enabling access to heterogeneous strain fields that encode constitutive behavior beyond what is available from global measurements \cite{hild2006digital,sutton2009image,grediac2012full,jones2018good,wei2026machine}. Infrared thermography provides complementary surface-temperature information, giving access to the thermal response associated with deformation, heat transfer, and thermomechanical coupling \cite{pitarresi2003review,rose2020optimisation}. When combined with global measurements such as reaction forces, these data sources provide complementary constraints that can improve the identifiability of coupled thermomechanical parameters \cite{rose2020optimisation,rose2021identification}.
However, these approaches also introduce additional challenges. In practice, measurements are typically restricted to surfaces, leading to incomplete observability of internal fields and necessitating careful formulation of inverse problems \cite{biegler2003large,seidl2022calibration}. Furthermore, thermal data are often noisy and of lower spatial resolution than displacement measurements, and differences in configurations (Eulerian versus Lagrangian) complicate data assimilation \cite{avril2008overview}. These challenges highlight the need for robust, physics-consistent inverse frameworks capable of handling finite strains, incompressibility, multiphysics coupling, and partial, noisy full-field data within a unified setting \cite{farrell2013automated,seidl2022calibration}.
In this work, we develop a physics-consistent framework for identifying thermomechanical material behavior from surface measurements. We formulate a fully coupled, PDE-constrained inverse problem that can incorporate displacement data from DIC, temperature fields from infrared thermography, and global reaction-force measurements. These data enter as weighted observation terms in the objective functional, while the known mechanical and thermal loading conditions are imposed through physical boundary conditions. The underlying forward model is based on a finite-strain, nearly incompressible thermo-hyperelastic formulation using a mixed $( \boldsymbol{u},p,\theta)$ description, enabling the treatment of large deformations relevant for soft materials and biological tissues. Thermomechanical coupling is derived in a thermodynamically consistent manner from a unified Helmholtz free energy, thereby ensuring a physically meaningful interaction between the mechanical response and the temperature evolution.
To assess the proposed framework, we consider both synthetically generated data and data obtained from experiments. In the synthetic setting, a forward problem is used to generate reference solutions under controlled conditions, while in the experimental setting, measured data are incorporated through objective terms in the inverse formulation. The resulting inverse problems are solved using adjoint-based gradients for a transient, nonlinear multiphysics system, enabling efficient optimization despite the high dimensionality of the problem. 

The main contributions of this work are fourfold. First, we formulate a finite-strain thermomechanical calibration problem that combines near-incompressible thermo-hyperelasticity with boundary force and surface displacement and temperature observations. Second, we separate physical loading conditions from observation data, allowing known mechanical and thermal inputs to enter as boundary conditions while noisy or incomplete measurements enter through weighted objective terms and, when appropriate, weak thermal assimilation. Third, we compute reduced gradients using adjoint sensitivities generated by automatic differentiation and verify them by finite differences. Fourth, we demonstrate the workflow on controlled synthetic full-field examples and on an experimental dataset in which thermomechanical shrinkage parameters are inferred from visible surface thermography and boundary-force histories.

Overall, the proposed framework enables data assimilation without requiring volumetric observations and extends inverse identification to fully coupled thermomechanics at finite strain. This provides a unified approach for extracting material parameters from partial and noisy full-field measurements while maintaining physical consistency and robustness.
The remainder of this paper is organized as follows. In Section~\ref{sec:Formulation}, we introduce the fully coupled thermomechanical formulation, including the governing equations, constitutive assumptions, and boundary conditions. Section~\ref{sec:inverse} presents the inverse problem, detailing the construction of the objective functional, the incorporation of measurement data, and the adjoint-based solution strategy. In Section~\ref{sec:results}, we demonstrate the performance of the proposed framework using both synthetically generated data and experimental measurements. Section~\ref{sec:conclusions} summarizes the main conclusions, limitations, and future directions.

\section{Thermomechanical Formulation}\label{sec:Formulation}
This section introduces the fully coupled thermomechanical formulation underlying the proposed framework. We consider a finite-strain continuum description in which the mechanical deformation and temperature evolution are treated as primary fields. The problem is posed on a reference configuration, and the governing equations are formulated in a manner consistent with thermodynamic principles. We begin by defining the kinematics and the associated problem setting, which provide the foundation for the subsequent balance laws and constitutive modeling.

\subsection{Kinematics and Problem Setting}
We consider a deformable body occupying the reference configuration $\Omega_0 \subset \mathbb{R}^3$ with boundary $\partial \Omega_0$. Material points are identified by their reference coordinates $\mathbf{X} \in \Omega_0$. The boundary is decomposed into disjoint parts associated with mechanical constraints, thermal loading, environmental interaction, and measurement surfaces, which are specified below.
The primary unknown fields of the problem are the displacement field $\mathbf{u} : \Omega_0 \rightarrow \mathbb{R}^3$, the pressure-like field $p : \Omega_0 \rightarrow \mathbb{R}$ enforcing near-incompressibility, and the temperature field $\theta : \Omega_0 \rightarrow \mathbb{R}$. The motion of the body is described by the mapping
$
\mathbf{x}(\mathbf{X},t) = \mathbf{X} + \mathbf{u}(\mathbf{X},t)$,
where $\mathbf{x}$ denotes the current position.
Finite deformations are characterized by the deformation gradient
\begin{equation}
\mathbf{F} = \mathbf{I} + \nabla \mathbf{u},
\end{equation}
where $\nabla$ denotes the gradient with respect to the reference coordinate $\mathbf{X}$ and $\mathbf{I}$ is the identity tensor. From $\mathbf{F}$, we define the right Cauchy--Green tensor and its determinant as
\begin{equation}
\mathbf{C} = \mathbf{F}^{\top} \mathbf{F}, 
\qquad
J = \det \mathbf{F}.
\end{equation}
To separate volumetric and isochoric contributions, we introduce the modified tensor
\begin{equation}
\bar{\mathbf{C}} = J^{-2/3} \mathbf{C}.
\end{equation}
The boundary $\partial \Omega_0$ is decomposed into disjoint regions associated with different types of mechanical and thermal boundary conditions. This partitioning is made problem-specific.

\subsection{Thermo-Hyperelastic Constitutive Model}

We model the material as a finite-strain thermo-hyperelastic solid characterized by a Helmholtz free energy density per unit reference volume \cite{miehe1995entropic}, denoted by $\Psi$, that depends on both deformation and temperature. The free energy is expressed as
\begin{equation}\label{eq:free_energy_general}
\Psi = \Psi(\mathbf{C}, \theta; \boldsymbol{m}),
\end{equation}
where $\mathbf{C}$ is the right Cauchy--Green deformation tensor, $\theta$ denotes the absolute temperature, and $\boldsymbol{m}$ is a set of material parameters. Equation~\eqref{eq:free_energy_general} provides the common constitutive starting point for the synthetic and experimental model specializations used below. To account for near-incompressibility, the free energy is decomposed into isochoric and volumetric contributions,
\begin{equation}\label{eq:free_energy_split}
\Psi(\mathbf{C}, \theta; \boldsymbol{m}) 
= \Psi_{\text{iso}}(\bar{\mathbf{C}}, \theta; \boldsymbol{m}) 
+ \Psi_{\text{vol}}(J, \theta; \boldsymbol{m}),
\end{equation}
where $\bar{\mathbf{C}}$ denotes the isochoric part of $\mathbf{C}$ and $J = \det \mathbf{F}$.
The volumetric contribution governs the compressibility of the material and incorporates thermal expansion or shrinkage effects through a temperature-dependent volumetric response. We write this response in terms of a thermal volumetric logarithmic strain $\varepsilon_{\mathrm{th}}^v(\theta;\boldsymbol{m})$ \cite{lu1975decomposition,lubarda2004constitutive}. For an isotropic thermal expansion model, $\varepsilon_{\mathrm{th}}^v = 3\alpha(\theta-\theta_0)$, while more general choices may include thermally activated shrinkage or anisotropic thermal distortion. This formulation enables the consistent treatment of nearly incompressible materials while accounting for thermally induced volume changes. The associated pressure-like variable arises naturally from this volumetric response and is used to enforce the incompressibility constraint.
The isochoric contribution describes the deviatoric mechanical behavior and captures the nonlinear response of soft materials under finite deformation. In particular, the shear response is allowed to depend on temperature, enabling the model to represent temperature-induced changes in stiffness commonly observed in polymers and biological tissues.
\subsection{Governing Equations}

We consider a fully coupled thermomechanical problem at finite strain in the reference configuration $\Omega_0$ \cite{miehe1995entropic}. The loading process is assumed to be quasi-static from a mechanical standpoint, so that inertial effects are neglected, and the mechanical problem is represented by a sequence of equilibrium states. Time dependence, therefore, enters the formulation through the thermal field and through the loading history only. The governing equations consist of the balance of linear momentum, a near-incompressibility constraint, and the balance of energy.

In the absence of body forces, the balance of linear momentum in the reference configuration is
\begin{equation}
\operatorname{Div}\,\mathbf{P} = \mathbf{0}
\qquad \text{in } \Omega_0,
\end{equation}
where $\mathbf{P}$ is the first Piola stress tensor. To account for the nearly incompressible response of soft materials, we introduce a pressure-like field $p$ and enforce a constitutive volumetric constraint $g(J,\theta,p)$, here chosen as
\begin{equation}\label{eq1:g}
g(J,\theta,p)
=
\frac{p}{K}
-
\left[
\ln J - \varepsilon_{\mathrm{th}}^v(\theta;\boldsymbol{m})
\right]
=0
\qquad \text{in } \Omega_0,
\end{equation}
where $K$ is a prescribed bulk penalty modulus, with units of stress, used to enforce near-incompressibility in the coupled mixed formulation, and $\varepsilon_{\mathrm{th}}^v$ denotes the prescribed temperature-dependent volumetric thermal strain.

The thermal field is governed by the balance of energy in referential form, where a superposed dot denotes the material time derivative,
\begin{equation}
\dot{\mathcal{E}} = \mathbf{P} : \dot{\mathbf{F}} - \operatorname{Div}\,\mathbf{Q} + R
\qquad \text{in } \Omega_0,
\end{equation}
where $\mathcal{E}$ is the internal energy density per unit reference volume, $\mathbf{Q}$ is the referential heat flux, and $R$ is a volumetric heat source per unit reference volume. The second law of thermodynamics is expressed through the Clausius--Duhem inequality,
\begin{equation}
\dot{\mathcal{H}} \geq - \operatorname{Div}\!\left(\frac{\mathbf{Q}}{\theta}\right) + \frac{R}{\theta},
\end{equation}
where $\mathcal{H}$ denotes the entropy density per unit reference volume.

Introducing the Helmholtz free energy density per unit reference volume,
\begin{equation}
\Psi = \mathcal{E} - \theta \mathcal{H},
\end{equation}
its material time derivative is
\begin{equation}
\dot{\Psi} = \dot{\mathcal{E}} - \dot{\theta}\,\mathcal{H} - \theta \dot{\mathcal{H}}.
\end{equation}
Substituting the balance of energy into this relation and using the Clausius--Duhem inequality yields
\begin{equation}
- \left( \dot{\Psi} + \mathcal{H} \dot{\theta} \right)
+ \mathbf{P} : \dot{\mathbf{F}}
- \frac{1}{\theta}\mathbf{Q}\cdot \nabla\theta
\geq 0.
\end{equation}

Introducing the second Piola--Kirchhoff stress $\mathbf{S}=\mathbf{F}^{-1}\mathbf{P}$, the dissipation inequality becomes
\begin{equation}
\left(
\mathbf{S} - 2 \frac{\partial \Psi}{\partial \mathbf{C}}
\right) : \frac{1}{2}\dot{\mathbf{C}}
-
\left(
\mathcal{H} + \frac{\partial \Psi}{\partial \theta}
\right)\dot{\theta}
-
\frac{1}{\theta}\mathbf{Q}\cdot \nabla\theta
\geq 0.
\end{equation}

Following the Coleman--Noll procedure \cite{coleman1974thermodynamics}, the rates $\dot{\mathbf{C}}$ and $\dot{\theta}$ are arbitrary, which yields
\begin{equation}
\mathbf{S} = 2 \frac{\partial \Psi}{\partial \mathbf{C}},
\qquad
\mathcal{H} = - \frac{\partial \Psi}{\partial \theta}.
\end{equation}
The remaining dissipation inequality reduces to the thermal restriction
\begin{equation}
- \frac{1}{\theta}\mathbf{Q}\cdot \nabla\theta \geq 0,
\end{equation}
which is satisfied, for example, by a Fourier-type heat flux law.

\subsection{Thermomechanical Coupling and Heat Flux}

We now make explicit the coupling between mechanical deformation and temperature evolution. Starting from the balance of energy and the thermodynamic relations derived in the previous section, the temperature evolution equation can be written as
\begin{equation}
c_\theta \dot{\theta} = \theta \frac{1}{2} \frac{\partial \mathbf{S}}{\partial \theta} : \dot{\mathbf{C}} - \operatorname{Div}\,\mathbf{Q}.
\end{equation}
Here $c_\theta$ denotes the effective volumetric heat capacity. Depending on the unit convention, this coefficient may represent $\rho_0 c_v$ or a directly calibrated volumetric heat-capacity parameter.
To express this contribution compactly, we introduce the thermoelastic coupling tensor
\begin{equation}
\mathbf{M} := \frac{1}{2} \frac{\partial \mathbf{S}}{\partial \theta}.
\end{equation}
Substituting this definition into the energy balance yields
\begin{equation}
c_\theta \dot{\theta} = \theta \mathbf{M} : \dot{\mathbf{C}} - \operatorname{Div}\,\mathbf{Q}.
\end{equation}
The introduction of $\mathbf{M}$ provides a compact representation of the thermomechanical source term and highlights that the coupling is governed by the temperature sensitivity of the stress response. This form is particularly convenient for both analysis and numerical implementation.
This term captures the reversible conversion between mechanical work and thermal energy, and its magnitude depends on the temperature sensitivity of the stress response.
The coupling tensor $\mathbf{M}$ inherits contributions from both the isochoric and volumetric parts of the free energy. Consequently, thermomechanical coupling arises from both deviatoric deformation mechanisms and volumetric effects such as thermal expansion.
The heat flux is modeled through a Fourier-type constitutive law expressed in the reference configuration \cite{wriggers1992coupled,wcislo2023spatial}. Starting from an isotropic conductivity in the current configuration, the flux is pulled back to the reference configuration, yielding
\begin{equation}
\mathbf{Q} = - J\, k_{\mathrm{therm}}\, \mathbf{C}^{-1} \nabla \theta,
\end{equation}
where $k_{\mathrm{therm}}$ is the thermal conductivity. This form ensures frame invariance and consistency with finite-strain kinematics.
The resulting temperature evolution equation, therefore, consists of a storage term, a conductive term, and a reversible thermoelastic source term. Importantly, both the stress tensor $\mathbf{S}$ and the coupling tensor $\mathbf{M}$ are derived from the same Helmholtz free energy, ensuring thermodynamic consistency of the formulation.

\subsection{Boundary and Loading Conditions}

The boundary $\partial \Omega_0$ is decomposed into regions associated with mechanical constraints, known thermal loading, measurement surfaces, and  environmental interactions without direct measurements. Mechanical boundary conditions are imposed through prescribed kinematic constraints that eliminate rigid body motion and represent the experimental or synthetic loading protocol. When displacement measurements are available, they may be used either to prescribe boundary motion or as observation data in the inverse objective, depending on the setting considered.

The thermal problem includes boundary conditions for environmental heat exchange and known thermal loading. On boundaries exposed to the environment, a convective boundary condition is imposed in the form
\begin{equation}
\mathbf{Q} \cdot \mathbf{N}
=
h_{\mathrm{conv}} \, (\theta - \theta_{\infty}),
\end{equation}
where $\mathbf{N}$ is the outward unit normal in the reference configuration, $h_{\mathrm{conv}}$ is an effective convection coefficient, and $\theta_{\infty}$ denotes the ambient temperature.
Known thermal loading is represented on a boundary portion $\Gamma_{\mathrm{act}}$ through a localized Robin-type condition,
\begin{equation}
\mathbf{Q} \cdot \mathbf{N}
=
h_{\mathrm{act}}\,w_{\mathrm{act}}(\mathbf{X})
\left(\theta - \theta_{\mathrm{act}}\right),
\end{equation}
where $w_{\mathrm{act}}$ defines the spatial footprint of the thermal actuator, $h_{\mathrm{act}}$ is an effective thermal transfer coefficient, and $\theta_{\mathrm{act}}$ is the prescribed actuator temperature or setpoint. Specific choices of $\Gamma_{\mathrm{act}}$, $w_{\mathrm{act}}$, and $\theta_{\mathrm{act}}$ depend on the synthetic or experimental loading protocol. Temperature measurements from thermography are treated as boundary observations in the inverse problem rather than as volumetric data.

\subsection{Weak Formulation and Discretization}

We formulate the coupled thermomechanical problem in a variational setting. Let $(\delta \mathbf{u}, \delta p, \delta \theta)$ denote admissible test functions associated with the displacement $\mathbf{u}$, pressure $p$, and temperature $\theta$, respectively. All equations are posed in the reference configuration $\Omega_0$. The weak form of the balance of linear momentum is given by
\begin{equation}
\int_{\Omega_0} \mathbf{P} : \nabla \delta \mathbf{u} \, d\mathbf{X} = 0,
\end{equation}
where $\mathbf{P}$ is the first Piola stress tensor. To account for the nearly incompressible response of soft materials, we introduce a pressure-like field $p$ \cite{simo1991quasi,simo1985variational} and enforce the volumetric constraint $g(J,\theta,p)=0$ as defined in Eq. \eqref{eq1:g}. The corresponding weak form is
\begin{equation}
\int_{\Omega_0}
\left(
\frac{p}{K}
-
\left[
\ln J - \varepsilon_{\mathrm{th}}^v(\theta;\boldsymbol{m})
\right]
\right)
\delta p \, d\mathbf{X} = 0,
\end{equation}
which corresponds to the variational enforcement of the constraint \eqref{eq1:g}.
The weak form of the transient heat equation reads
\begin{equation}
\begin{aligned}
&\int_{\Omega_0} c_\theta \dot{\theta} \, \delta \theta \, d\mathbf{X}
- \int_{\Omega_0} \theta \mathbf{M} : \dot{\mathbf{C}} \, \delta \theta \, d\mathbf{X}
- \int_{\Omega_0} \mathbf{Q} \cdot \nabla \delta \theta \, d\mathbf{X}
\\
&\quad
+ \int_{\Gamma_{\mathrm{env}}} h_{\mathrm{conv}}(\theta-\theta_\infty)\delta\theta\,d\Gamma
+ \int_{\Gamma_{\mathrm{act}}} h_{\mathrm{act}}w_{\mathrm{act}}(\mathbf{X})(\theta-\theta_{\mathrm{act}})\delta\theta\,d\Gamma
= 0,
\end{aligned}
\end{equation}
where $\mathbf{M}$ is the thermoelastic coupling tensor, $\mathbf{Q}$ is the referential heat flux, and the boundary terms represent environmental convection and known thermal actuation. Let $t_n$, $n=0,\ldots,N$, denote the discrete time levels and let $\Delta t_n=t_n-t_{n-1}$. The thermal problem is discretized in time using a backward Euler scheme, such that
\begin{equation}
\dot{\theta} \approx \frac{\theta^n - \theta^{n-1}}{\Delta t_n},
\end{equation}
and the coupling term involving $\dot{\mathbf{C}}$ is evaluated consistently using deformation measures at consecutive time steps. The spatial discretization is based on a mixed finite element formulation with approximation spaces
\begin{equation}
\mathbf{u} \in [\mathcal{P}_2]^3, 
\qquad
p \in \mathcal{P}_1, 
\qquad
\theta \in \mathcal{P}_1,
\end{equation}
ensuring stability for nearly incompressible materials through a Taylor--Hood-type displacement--pressure pair \cite{simo1985variational,simo1991quasi,brink1996some}. 
Motivated by semi-implicit and staggered treatments used in coupled thermoelastic finite element schemes \cite{farhat1991unconditionally}, the thermoelastic source term is evaluated in a lagged history-variable form rather than fully implicitly in the current Newton linearization. Specifically, the continuous scalar source
\begin{equation}
s_{\mathrm{th}} = \theta\,\mathbf{M}:\dot{\mathbf{C}}
\end{equation}
is represented at the discrete level by an elementwise projected scalar history field. At time step $t_n$, with $\Delta t_n=t_n-t_{n-1}$, the thermal residual uses the stored source value $s_{\mathrm{th}}^{n-1}$,
\begin{equation}
\begin{aligned}
&\int_{\Omega_0} c_\theta
\frac{\theta^n-\theta^{n-1}}{\Delta t_n}\,\delta\theta\,d\mathbf{X}
-
\int_{\Omega_0} s_{\mathrm{th}}^{n-1}\delta\theta\,d\mathbf{X}
-
\int_{\Omega_0}\mathbf{Q}^n\cdot\nabla\delta\theta\,d\mathbf{X}
+\cdots
=0 .
\end{aligned}
\end{equation}
The source history is held fixed during the nonlinear solve for time step $n$. After the coupled solve has converged, it is updated from the accepted increment according to
\begin{equation}
s_{\mathrm{th}}^{n}
=
\Pi_{\mathrm{DG0}}
\left[
\theta^{n-1}
\mathbf{M}^{n-1/2}:
\frac{\mathbf{C}^{n}-\mathbf{C}^{n-1}}{\Delta t_n}
\right],
\qquad
\mathbf{M}^{n-1/2}
=
\frac{1}{2}
\left(
\mathbf{M}^{n}+\mathbf{M}^{n-1}
\right),
\end{equation}
where $\Pi_{\mathrm{DG0}}$ denotes the $L^2$ projection onto elementwise constant fields. For the first time step, the source history is initialized as
$s_{\mathrm{th}}^{0}=0$, corresponding to an initially equilibrated state with no prior deformation increment. This treatment retains an increment-based approximation of reversible thermoelastic heating/cooling while reducing the strongest direct nonlinear feedback from the Newton solve. In the inverse calculations, this same history update is recorded in the automatic-differentiation tape, so the reported adjoint gradients correspond to the stabilized discrete forward model rather than to a different fully implicit source discretization.
The resulting discretization yields a coupled nonlinear system for the unknown fields $(\mathbf{u}, p, \theta)$ at each time step, which is solved in a monolithic manner to ensure a consistent treatment of the thermomechanical coupling.

\section{Inverse Model}\label{sec:inverse}

\subsection{Parameter Identification Problem}

We formulate the parameter identification problem as a partial differential equation (PDE)-constrained optimization problem \cite{biegler2003large,seidl2022calibration}. The unknown material parameters are collected in a vector
\begin{equation}
\boldsymbol{m} \in \mathcal{M},
\end{equation}
where $\mathcal{M}$ denotes the admissible parameter space. In practice, $\mathcal{M}$ may include bound constraints and prior information reflecting physically meaningful parameter ranges. For a given parameter vector, the thermomechanical forward problem introduced in Section~\ref{sec:Formulation} is solved over the loading history to obtain the state variables $(\mathbf{u}^n,p^n,\theta^n)$ at each time step.

The inverse problem is defined by minimizing the mismatch between model predictions and experimental observations. The available observations may include displacement fields from DIC, surface temperature fields from infrared thermography, and global response quantities such as reaction-force histories. These boundary measurements  may be incomplete or unavailable for some portions of the loading history. Known mechanical and thermal loading conditions are imposed through the physical boundary conditions of the forward problem, while measured quantities enter the inverse problem as weighted observation terms.

\subsection{Objective Functional}

We define a weighted least-squares objective functional over the parameter vector $\boldsymbol{m}$. A representative form of the objective is
\begin{equation}
\begin{aligned}
\mathcal{J}(\boldsymbol{m})
=
\sum_{n=1}^{N}
\Bigg[
&\frac{w_u}{2}\int_{\Gamma_u}
\chi_u^n
\left|
\Pi_u
\left(
\mathbf{u}^n(\boldsymbol{m}) - \tilde{\mathbf{u}}^n
\right)
\right|^2 d\Gamma
+
\frac{w_\theta}{2}\int_{\Gamma_\theta}
\chi_\theta^n
\left( \theta^n(\boldsymbol{m}) - \tilde{\theta}^n \right)^2 d\Gamma \\
&+
\frac{w_F}{2}
\left|
\mathbf{f}_{\mathrm{rxn}}^n(\boldsymbol{m})
-
\tilde{\mathbf{f}}_{\mathrm{rxn}}^n
\right|^2
\Bigg]
+
\mathcal{R}(\boldsymbol{m}) .
\end{aligned}
\end{equation}
Here, the sum is taken over the discrete observation times introduced above. The fields $\tilde{\mathbf{u}}^n$ and $\tilde{\theta}^n$ denote measured displacement and temperature data at time step $n$, respectively, while $\tilde{\mathbf{f}}_{\mathrm{rxn}}^n$ denotes the measured reaction-force resultant. The surfaces $\Gamma_u$ and $\Gamma_\theta$ denote the displacement- and temperature-observation parts of the boundary. The masks $\chi_u^n$ and $\chi_\theta^n$ indicate where displacement and temperature measurements are available and valid at time step $n$. The operator $\Pi_u$ maps the model displacement to the measured displacement components, for example, by selecting the in-plane DIC components. The weights $w_u$, $w_\theta$, and $w_F$ control the relative influence of the different observation types, and $\mathcal{R}(\boldsymbol{m})$ denotes an optional regularization term used to incorporate prior information or penalize nonphysical parameter values.

The objective weights are chosen so that the active observation terms have comparable magnitudes at the initial inverse iterate. This scaling prevents one data type from dominating the early optimization solely because of its units or numerical magnitude, while still allowing the subsequent optimization trajectory to be determined by the coupled model response and the measurement misfits.
Terms whose corresponding data are unavailable are omitted by setting the associated weight to zero. This produces a flexible objective that can be specialized to synthetic data, experimental thermography, DIC measurements, reaction-force histories, or combinations of these data sources.

Let the discrete state at time step $n$ be denoted by
\begin{equation}
\mathbf{y}^n =
\left(
\mathbf{u}^n,
p^n,
\theta^n
\right),
\end{equation}
where lagged or internal history variables are included in $\mathbf{y}^n$ when required by the time-discrete formulation. The fully discrete thermomechanical residual is written as
\begin{equation}
\mathbf{R}_n
\left(
\mathbf{y}^n,
\mathbf{y}^{n-1};
\boldsymbol{m}
\right)
=
\mathbf{0},
\qquad n=1,\ldots,N.
\end{equation}
In the present monolithic formulation, this residual contains the discrete mechanical equilibrium equation, the pressure constraint, and the transient thermal balance,
\begin{equation*}
\mathbf{R}_n
=
\begin{bmatrix}
\mathbf{R}_{u,n}\\
\mathbf{R}_{p,n}\\
\mathbf{R}_{\theta,n}
\end{bmatrix}
=
\mathbf{0}.
\end{equation*}
Here, $\mathbf{R}_{u,n}$ corresponds to the weak form of momentum balance, $\mathbf{R}_{p,n}$ to the near-incompressibility constraint, and $\mathbf{R}_{\theta,n}$ to the heat equation including thermoelastic coupling and thermal boundary terms. The dependence on $\mathbf{y}^{n-1}$ represents the backward-Euler time discretization and the history dependence introduced by the thermoelastic coupling term.
The inverse problem can therefore be written in reduced PDE-constrained form as
\begin{equation}
\begin{aligned}
\min_{\boldsymbol{m}\in\mathcal{M}} \quad
& \mathcal{J}(\boldsymbol{m}) \\
\text{s.t.} \quad
& \mathbf{R}_n(\mathbf{y}^n,\mathbf{y}^{n-1};\boldsymbol{m}) = \mathbf{0},
\qquad n = 1,\ldots,N,
\end{aligned}
\end{equation}
where $\mathbf{R}_n$ denotes the discrete weak form of the coupled thermomechanical governing equations at time step $n$.

\subsection{Adjoint Sensitivities and Optimization}

The solution of the inverse problem requires gradients of the objective functional with respect to the material parameters. A direct finite-difference evaluation would require repeated forward solves for each parameter and becomes computationally expensive for transient, nonlinear multiphysics problems. We therefore compute sensitivities using an adjoint-based approach in combination with automatic differentiation \cite{farrell2013automated}.

The objective can be written as a sum of per-step observation terms,
\begin{equation}
\mathcal{J}(\boldsymbol{m})
=
\sum_{n=1}^{N}
\Phi_n(\mathbf{y}^n;\boldsymbol{m})
+
\mathcal{R}(\boldsymbol{m}),
\end{equation}
where $\Phi_n$ contains the displacement, temperature, and reaction-force misfit terms at time step $n$.
To derive the discrete adjoint equations, we introduce adjoint variables $\boldsymbol{\lambda}^n$ associated with the residual constraints and define the Lagrangian
\begin{equation}
\begin{aligned}
\mathcal{L}
=
\sum_{n=1}^{N}
\Phi_n(\mathbf{y}^n;\boldsymbol{m})
+
\mathcal{R}(\boldsymbol{m})
+
\sum_{n=1}^{N}
(\boldsymbol{\lambda}^n)^{\top}
\mathbf{R}_n
\left(
\mathbf{y}^n,
\mathbf{y}^{n-1};
\boldsymbol{m}
\right).
\end{aligned}
\end{equation}
Stationarity of $\mathcal{L}$ with respect to the state variables gives a terminal adjoint problem that is solved backward in time. For the final time step,
\begin{equation}
\left(
\frac{\partial \mathbf{R}_N}{\partial \mathbf{y}^N}
\right)^{\top}
\boldsymbol{\lambda}^N
=
-
\left(
\frac{\partial \Phi_N}{\partial \mathbf{y}^N}
\right)^{\top},
\end{equation}
and for $n=N-1,\ldots,1$,
\begin{equation}
\begin{aligned}
\left(
\frac{\partial \mathbf{R}_n}{\partial \mathbf{y}^n}
\right)^{\top}
\boldsymbol{\lambda}^n
=
&-
\left(
\frac{\partial \Phi_n}{\partial \mathbf{y}^n}
\right)^{\top} \\
&-
\left(
\frac{\partial \mathbf{R}_{n+1}}{\partial \mathbf{y}^n}
\right)^{\top}
\boldsymbol{\lambda}^{n+1}.
\end{aligned}
\end{equation}
After the adjoint variables have been computed, the reduced gradient of the objective with respect to the material parameters is obtained as
\begin{equation}
\frac{d\mathcal{J}}{d\boldsymbol{m}}
=
\frac{\partial \mathcal{R}}{\partial \boldsymbol{m}}
+
\sum_{n=1}^{N}
\left[
\frac{\partial \Phi_n}{\partial \boldsymbol{m}}
+
(\boldsymbol{\lambda}^n)^{\top}
\frac{\partial \mathbf{R}_n}{\partial \boldsymbol{m}}
\right].
\end{equation}
This expression avoids finite-difference approximations and requires only one forward solution and one backward adjoint solution for each gradient evaluation. The adjoint equations above are written at the monolithic residual level. Equivalently, the operators $\partial \mathbf{R}_n/\partial \mathbf{y}^n$ and $\partial \mathbf{R}_{n+1}/\partial \mathbf{y}^n$ contain the coupled mechanical, constraint, and thermal Jacobian blocks implied by the block residual above. Automatic differentiation is used to evaluate these transposed Jacobian actions and the parameter derivatives consistently with the nonlinear finite element implementation, rather than deriving each block by hand \cite{farrell2013automated,seidl2022calibration}. This is analogous in spirit to more detailed global--local adjoint derivations used for path-dependent constitutive calibration \cite{seidl2022calibration}, but here the residual is treated as a coupled thermomechanical finite element residual.

The resulting reduced optimization problem is solved using adjoint-based gradients in a scaled parameter space. Each material parameter is non-dimensionalized by a prior or reference value so that the optimization variables have comparable magnitudes. Within this reduced-space formulation, we consider two gradient-based update strategies. The first is a bound-constrained quasi-Newton method using L-BFGS-B \cite{liu1989limited}, which uses objective and adjoint-gradient evaluations to build an approximate inverse Hessian. The second is a scaled gradient-descent method with backtracking line search, which provides a simpler fallback for cases in which quasi-Newton updates lead to failed nonlinear solves or unsuitable trial parameters. In both cases, the governing thermomechanical equations are solved at each trial parameter value, and the adjoint gradient is computed with respect to the selected material parameters.

\section{Results}\label{sec:results}

This section evaluates the proposed inverse thermomechanical framework in progressively less controlled settings. We first consider synthetic data generated with the same finite element model, where the ground-truth material parameters and loading histories are known. These examples are used to assess whether the selected observations, including surface displacement, surface temperature, and global reaction forces, contain sufficient information to recover the parameters of interest. For all inverse studies, including the experimental thermal calibration, the corresponding finite-difference checks of the adjoint gradients are reported separately in Appendix~\ref{app:fd_checks}. We then apply the same inverse modeling workflow to experimental measurements, where the mechanical response is calibrated from equibiaxial loading data and the thermal inverse problem is driven by measured temperature and force histories.
For visualization, objective histories are plotted as $J_i/J_0$, where $J_0$ is the total objective value at the initial inverse iterate. For plots containing multiple objective components, each component is divided by the same $J_0$, so the curves show their contributions to the normalized total objective.

The synthetic studies are designed to separate different aspects of the coupled inverse problem. The first example uses a spatially uniform thermal preconditioning stage followed by displacement-controlled loading, providing a controlled setting for evaluating the coupling between temperature, deformation, and reaction force. The second example introduces localized transient heating through repeated rod contact during staged mechanical loading, which produces a more heterogeneous thermomechanical response and more closely reflects the type of boundary-driven thermal actuation encountered in experiments.

\subsection{Synthetic Model Setup}

The synthetic examples use a common three-dimensional plate geometry to generate controlled thermomechanical data with known material parameters and loading histories. The reference domain is a rectangular plate,
\begin{equation}
\Omega_0 \subset [0,L_0] \times [0,L_0] \times [0,T_0],
\end{equation}
where $L_0 = 100~\mathrm{mm}$ and $T_0 = 10~\mathrm{mm}$. Two through-thickness elliptical holes are removed from the plate, producing heterogeneous deformation and temperature fields even under nominally simple thermal and mechanical boundary conditions. The geometry and applied boundary conditions are shown with the individual synthetic problem setups below.

All synthetic data are generated by solving the coupled thermomechanical finite element problem described in Section~\ref{sec:Formulation}. The displacement, pressure-like, and temperature fields are approximated using the mixed finite element discretization described previously. Synthetic observations are then extracted from selected boundary surfaces and from the reaction force on the loaded boundary, matching the data types used in the inverse objective. The DIC-like displacement observations are obtained from the in-plane displacement components on the top surface. Because the measurement points and the inverse finite element surface nodes may not coincide, the measured surface fields are interpolated onto the observed boundary of the inverse mesh before the objective is evaluated. For each inverse-mesh surface node, the measured field is evaluated at the same physical surface coordinate using piecewise-linear interpolation over a triangulation of the measurement points on the planar surface; nearest-neighbor assignment is used only for points outside the triangulated measurement domain. The DIC-like displacement data are used in two distinct ways. A scalar axial displacement history is extracted from points near the loaded boundary at $x=L_0=100~\mathrm{mm}$ and prescribed as the displacement-control boundary condition on that face. The full-field top-surface displacement data are retained separately as observations in the inverse objective. The opposite face at $x=0$ is fully clamped, the transverse displacement components on the loaded $x=L_0$ face are not constrained, and the remaining exterior faces are mechanically traction-free. Thus, the bottom surface is not mechanically fixed and is not assigned DIC displacement data.
Termography-like temperature observations are treated analogously as scalar surface fields on the visible top surface or on the localized bottom measurement region.
For the synthetic examples, the isochoric response is represented by a temperature-dependent finite-chain-corrected neo-Hookean model,
\begin{equation}
\Psi_{\mathrm{iso}}(\bar{\mathbf{C}},\theta)
=
\frac{G(\theta)}{2}\left( \mathrm{tr}(\bar{\mathbf{C}}) - 3 \right),
\qquad
G(\theta) = N_R k_B \theta \, \zeta, \quad \text{with} \quad N_R = \frac{G_{\mathrm{shear},0}}{k_B \theta_0} \, .
\end{equation}
The function $\zeta$ represents a finite-chain extensibility correction and is defined in terms of the normalized isochoric stretch as
\begin{equation}
z = \frac{\bar{\lambda}}{\lambda_L},
\end{equation}
where $\bar{\lambda} = \sqrt{\mathrm{tr}(\bar{\mathbf{C}})/3}$ and $\lambda_L$ denotes the limiting chain extensibility. 
The correction factor is then given by
\begin{equation}
\zeta =
\frac{\lambda_L}{3 \bar{\lambda}}
\frac{z(3 - z^2)}{1 - z^2},
\end{equation}
which captures the nonlinear stiffening due to finite chain extensibility.
The following values define the baseline synthetic material model unless otherwise stated; $\lambda_L$ is dimensionless:
\begin{equation}
G_{\mathrm{shear},0} = 280~\mathrm{kPa},
\qquad
\lambda_L = 5.12,
\qquad
k_B = 1.38 \times 10^{-17}~\mathrm{kPa\,mm^3/K},
\qquad
\theta_0 = 293~\mathrm{K}.
\end{equation}

The volumetric response incorporates thermal expansion through a penalty-type energy
\begin{equation}
\Psi_{\mathrm{vol}}(J,\theta)
=
\frac{K_{\mathrm{bulk}}}{2}
\left( \ln J - 3\alpha(\theta - \theta_0) \right)^2,
\end{equation}
which enforces near-incompressibility while allowing for temperature-dependent volumetric deformation. The bulk modulus is chosen as
\begin{equation}
K_{\mathrm{bulk}} = 1000\, G_{\mathrm{shear},0},
\end{equation}
so that it acts as a numerical near-incompressibility penalty rather than an independently identified material parameter. The thermal expansion coefficient is set to
\begin{equation}
\alpha = 180 \times 10^{-6}~\mathrm{K}^{-1}.
\end{equation}
The remaining thermal constants, ground-truth values, and inversion controls are specified with each synthetic problem. The individual examples differ in the thermal actuation protocol, the observation fields included in the objective, and the subset of material parameters treated as inversion controls.

In both synthetic examples, the initial inverse guess is obtained by scaling the identified ground-truth parameters by $1.1$. This choice is used only to provide a reproducible non-exact starting point for the optimizer; the specific $10\%$ perturbation has no special significance. For each synthetic inverse problem, the active objective weights are selected from the initial objective breakdown so that the displacement, temperature, and force contributions are approximately balanced at the starting point. The synthetic data are generated without added measurement noise or model-form error. This is intentional: the purpose of the synthetic studies is to isolate the internal consistency and parameter recoverability of the proposed inverse formulation under controlled conditions. The effect of experimental noise, incomplete measurements, and model discrepancy is examined subsequently through the experimental dataset.

\subsubsection{Synthetic Problem 1: Uniform Thermal Preconditioning}

\begin{figure}[htbp]
    \centering
    \includegraphics[width=0.5\linewidth]{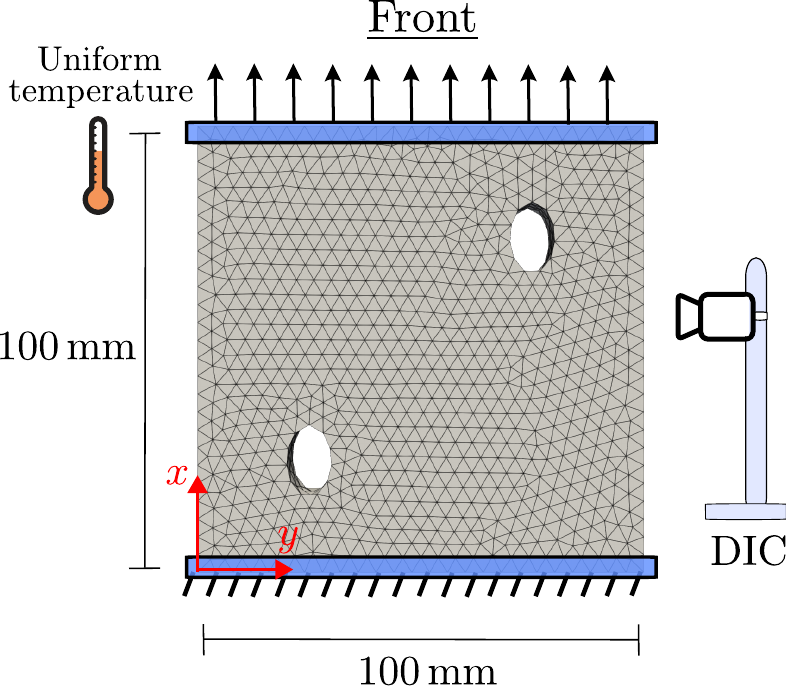}
    \caption{Synthetic uniform-preconditioning setup. The perforated plate is exposed to a spatially uniform chamber temperature while mechanically fixed, and is then pulled in displacement control on the $x=L_0=100~\mathrm{mm}$ face while the $x=0$ face remains clamped. The observable surface displacement field and the reaction force on the loaded face define the mechanical observations used in the inverse problem.}
    \label{fig:syn1_setup}
\end{figure}

The first synthetic problem is a uniform chamber-heating experiment followed by displacement-controlled uniaxial loading. It provides a controlled setting in which the thermal state is established before mechanical loading begins, allowing the influence of thermoelastic material parameters on the subsequent force-displacement response to be isolated. The setup is shown schematically in Fig.~\ref{fig:syn1_setup}.

The specimen is initially at the ambient reference temperature,
\begin{equation}
\theta(\mathbf{X},0)=\theta_0,
\qquad
\theta_0 = 293~\mathrm{K}.
\end{equation}
During the first stage, the specimen is exposed to a spatially uniform chamber-type Robin condition while the mechanical displacement is held fixed. The chamber temperature is a known thermal input, not an inversion variable. It is ramped by $\Delta\theta=100~\mathrm{K}$ from $293~\mathrm{K}$ to $393~\mathrm{K}$ over $20$ thermal steps with total pseudo-time $t_{\mathrm{therm}}=1.0$. The chamber temperature is then held fixed at $393~\mathrm{K}$ for $15$ additional thermal-dwell steps with total pseudo-time $t_{\mathrm{hold}}=5000.0$. This dwell allows the thermally preconditioned state to approach equilibrium before mechanical loading begins.

After the thermal dwell, the chamber temperature remains fixed at the elevated value and the specimen is pulled quasi-statically in displacement control. The $x=0$ face is fully clamped and the $x=L_0=100~\mathrm{mm}$ face is prescribed an axial displacement,
\begin{equation}
\mathbf{u}=\mathbf{0}
\quad \text{on } x=0,
\qquad
u_x=\bar{u}_x(t)
\quad \text{on } x=L_0 .
\end{equation}
The transverse displacement components on the pulled face are not constrained. The prescribed displacement $\bar{u}_x(t)$ is increased from $0$ to $20~\mathrm{mm}$ over $50$ mechanical loading steps with total pseudo-time $t_{\mathrm{mech}}=1.0$. This prescribed displacement defines the mechanical loading condition; it is not treated as a displacement observation in the inverse objective.

The thermal transport parameters are held fixed in this example and are chosen to define a controlled synthetic thermal response:
\begin{equation}
c_\theta = 1839~\mathrm{mN/(mm^2\,K)},
\qquad
k_{\mathrm{therm}} = 400~\mathrm{mN/(s\,K)},
\qquad
h_{\mathrm{contact}} = 1000~\mathrm{mN/(mm\,s\,K)} .
\end{equation}
The inverse controls are the ground-state shear modulus and the thermal expansion coefficient,
\begin{equation}
\boldsymbol{m}
=
\left[
G_{\mathrm{shear},0},
\alpha
\right].
\end{equation}
The inverse solve is initialized using the common perturbed starting point described above; explicitly,
\begin{equation}
G_{\mathrm{shear},0}^{(0)} = 1.1\,G_{\mathrm{shear},0}^{\mathrm{true}},
\qquad
\alpha^{(0)} = 1.1\,\alpha^{\mathrm{true}} .
\end{equation}

The synthetic observations consist of the in-plane displacement field on the observable top surface and the global reaction-force history on the loaded boundary. The DIC-like displacement data are used both as a full-field observation and to define the scalar loading coordinate for the force-displacement response. Specifically, the displacement data are reduced to an average axial displacement over the observed loaded-edge region near $x=L_0$. This average is used as the displacement coordinate for the force-displacement response, so the force curve is compared against the measured specimen motion rather than only the nominal prescribed displacement. Thus, the full-field term constrains the spatial deformation pattern, while the averaged edge displacement defines the scalar loading coordinate used in the mechanical response comparison. The temperature field is not included as a misfit term in this example, so $w_{\theta}=0$. The objective, therefore, combines the top-surface displacement-field mismatch, the DIC-derived average displacement-curve mismatch, and the reaction-force mismatch on the loaded $x=L_0$ boundary,
\begin{equation}
J = J_u + J_{\mathrm{disp}} + J_{\mathrm{force}} .
\end{equation}
This problem tests whether the coupled inverse formulation can recover thermoelastic parameters from the mechanically observed response after a known thermal preconditioning stage. The corresponding adjoint gradients were verified against finite-difference gradients, as reported in Appendix~\ref{app:fd_syn1}.

\begin{figure}[H]
    \centering
    \begin{subfigure}[b]{0.49\linewidth}
        \centering
        \includegraphics[width=\linewidth]{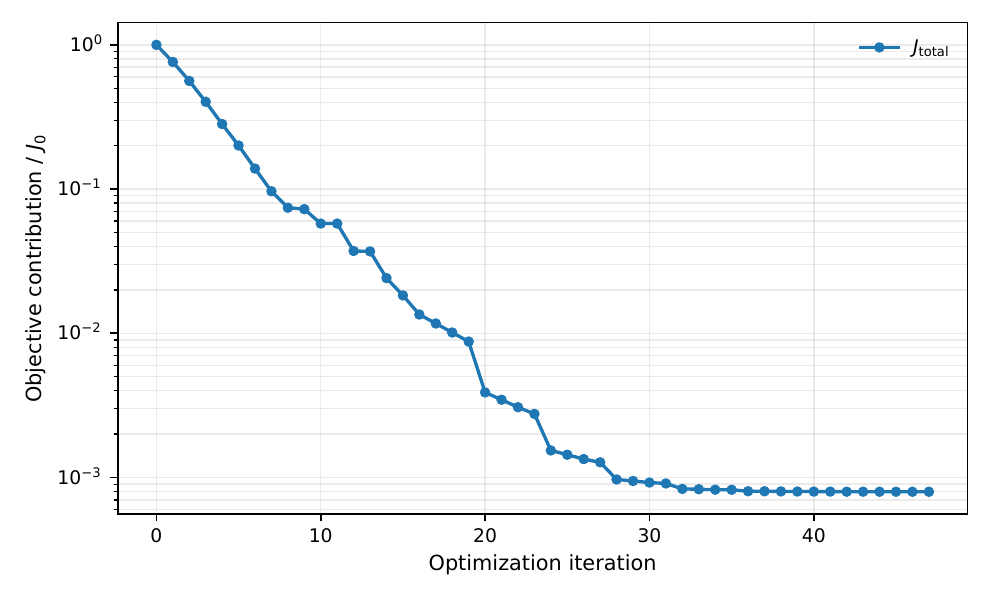}
        \caption{Objective convergence.}
    \end{subfigure}\hfill
    \begin{subfigure}[b]{0.49\linewidth}
        \centering
        \includegraphics[width=\linewidth]{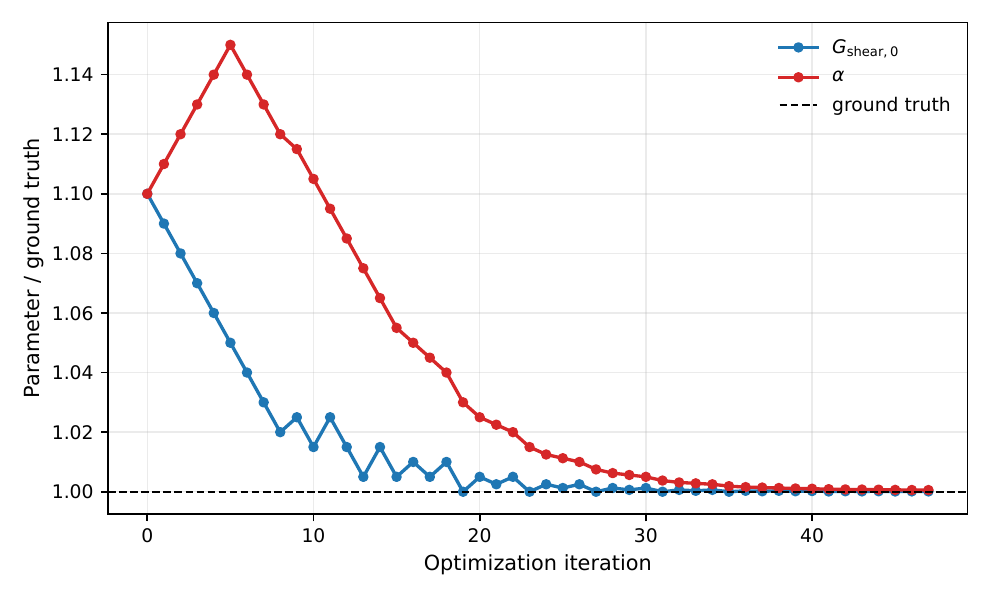}
        \caption{Parameter recovery.}
    \end{subfigure}

        \begin{subfigure}[b]{0.49\linewidth}
        \centering
        \includegraphics[width=\linewidth]{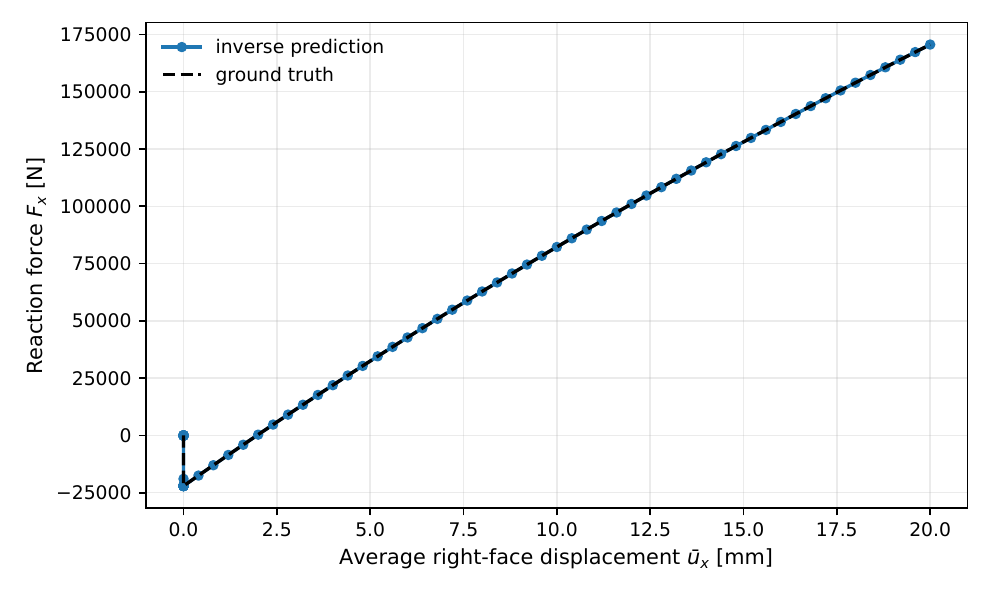}
        \caption{Force-displacement response.}
    \end{subfigure}

    \caption{Inverse results for the uniform thermal-preconditioning problem. The objective decreases over the optimization history, the identified parameters approach their ground-truth values, and the final force-displacement response agrees with the synthetic reference data.}
    \label{fig:syn1_inverse_results}
\end{figure}

Figure~\ref{fig:syn1_inverse_results} shows that the inverse problem recovers both active material parameters from the mechanically observed response. Starting from the common perturbed initial guess, the normalized objective decreases by more than three orders of magnitude before reaching a plateau. The recovered shear modulus is essentially indistinguishable from the ground truth, while the thermal expansion coefficient initially overshoots and then relaxes toward its true value. This behavior indicates that the thermal expansion coefficient is identifiable from the preconditioned mechanical response, but is more strongly coupled to the transient thermal history than the shear modulus. The final force-displacement curve agrees closely with the synthetic reference data over the full loading range. The nonzero force branch at zero applied displacement corresponds to the restrained thermal-preconditioning stage, during which thermal expansion generates reaction forces before mechanical pulling begins.
\subsubsection{Synthetic Problem 2: Staged Rod-Contact Loading}
\begin{figure}[H]
    \centering
    \includegraphics[width=1.0\linewidth]{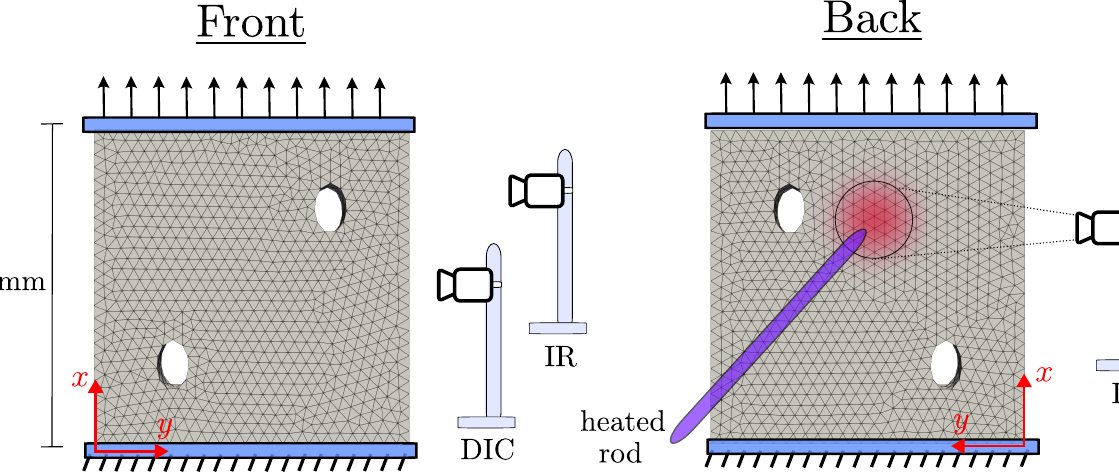}
    \caption{Synthetic staged rod-contact setup. A localized temperature-controlled contact region is applied on the bottom surface by a heated rod. The front-side measurements represent the DIC-like in-plane displacement field and an IR-like temperature field over the observable top surface. The back-side IR schematic indicates an additional local specimen-temperature observation near the rod-contact region, not a full-field back-side temperature measurement. During the inverse problem, the rod setpoint and contact timing are treated as known thermal inputs, while the measured top-surface and local bottom-surface temperatures enter only as observation terms.}
    \label{fig:syn2_setup}
\end{figure}

The second synthetic problem uses a staged thermomechanical protocol in which a temperature-controlled rod is repeatedly brought into contact with the bottom surface of the specimen. This problem is designed to test whether the inverse formulation can recover material parameters from transient, localized thermal actuation combined with the mechanical reaction-force response during piecewise displacement-controlled loading. The setup and loading sequence are shown in Fig.~\ref{fig:syn2_setup}.

The specimen is initially at the ambient reference temperature,
\begin{equation}
\theta(\mathbf{X},0)=\theta_0,
\qquad
\theta_0 = 293~\mathrm{K}.
\end{equation}
The rod is represented by a localized Gaussian-weighted thermal contact footprint on the bottom face. The footprint has width parameter $\sigma=10~\mathrm{mm}$ and is centered at $(x,y)=(75,50)~\mathrm{mm}$. During active contact, the rod/controller setpoint is
\begin{equation}
\theta_{\mathrm{rod}} = \theta_0 + 100~\mathrm{K} = 393~\mathrm{K}.
\end{equation}
When the rod is not in contact, the localized contact term is deactivated and the specimen cools through the ambient-convection boundaries.

The loading history begins with an ambient pre-contact segment at zero displacement. This segment uses $12$ steps over $45~\mathrm{s}$ and provides a force baseline before thermal actuation. The subsequent protocol consists of three rod-contact cycles. In each cycle, the displacement is held fixed while the rod is applied for $12$ heating steps over $30~\mathrm{s}$, the rod is removed for $12$ cooling steps over $45~\mathrm{s}$, and the specimen is then pulled to the next displacement level using six quasi-static loading increments over $1~\mathrm{s}$.

The prescribed displacement levels are
\begin{equation}
\bar{u}_x \in
\{0,\ 0.33u_{\max},\ 0.66u_{\max},\ u_{\max}\},
\qquad
u_{\max}=7.0~\mathrm{mm},
\end{equation}
or equivalently
\begin{equation}
\bar{u}_x \in
\{0,\ 2.31,\ 4.62,\ 7.0\}~\mathrm{mm}.
\end{equation}
Thus, the rod first contacts the undeformed specimen, the specimen is pulled to $2.31~\mathrm{mm}$, the second rod contact is applied at that held displacement, the specimen is pulled to $4.62~\mathrm{mm}$, the third rod contact is applied at that held displacement, and the final pull reaches $7.0~\mathrm{mm}$.

The mechanical boundary conditions are
\begin{equation}
\mathbf{u}=\mathbf{0}
\quad \text{on } x=0,
\qquad
u_x=\bar{u}_x(t)
\quad \text{on } x=L_0 .
\end{equation}
The transverse displacement components on the pulled face are not constrained. In the inverse problem, the displacement history on the loaded $x=L_0$ face is not treated as an independent actuator measurement. Instead, it is obtained from the DIC-like top-surface displacement data by projecting or averaging the measured axial displacement near the loaded $x=L_0$ boundary, and the resulting displacement history is imposed as the mechanical Dirichlet input. The measured top-surface displacement field itself remains an observation in the objective.

The thermal boundary conditions in the inverse model are kept separate from the temperature observations. The rod/controller setpoint and the contact timing are treated as known inputs, analogous to a controlled heating device. During contact, the bottom surface is driven by the localized Robin contact law
\begin{equation}
-k_{\mathrm{therm}}\nabla\theta\cdot\mathbf{N}
=
h_{\mathrm{contact}}\,c_{\mathrm{rod}}^n\,w_{\mathrm{rod}}(\mathbf{X})
\left(\theta-\theta_{\mathrm{rod}}\right)
\quad \text{on } \Gamma_{\mathrm{bottom}},
\end{equation}
where $c_{\mathrm{rod}}^n$ is the time-dependent contact indicator and $w_{\mathrm{rod}}$ is the Gaussian footprint. The contact term is deactivated when the rod is lifted. The top surface and the remaining exposed surfaces are modeled with ambient-convection Robin conditions; the measured top-surface and bottom near-rod temperatures are not used as thermal reservoirs in the PDE.

The synthetic observations consist of four measurement streams. The first is the DIC-like in-plane displacement field on the top surface. The second is the top-surface temperature field, representing an IR-like surface temperature measurement. The third is a local bottom-surface specimen-temperature measurement near the rod-contact region; this field is used only as an observation and is not imposed as a thermal boundary condition. The fourth is the global reaction-force history on the loaded boundary. The resulting inverse objective is
\begin{equation}
\begin{aligned}
J
=
\sum_{n=1}^{N}
\Bigg[
&\frac{w_u}{2}\int_{\Gamma_{\mathrm{top}}}
\left|
\mathbf{u}_{\parallel}^n-\tilde{\mathbf{u}}_{\parallel}^n
\right|^2\,d\Gamma
+
\frac{w_{\theta,\mathrm{top}}}{2}\int_{\Gamma_{\mathrm{top}}}
\left(\theta^n-\tilde{\theta}_{\mathrm{top}}^n\right)^2\,d\Gamma
\\
&+
\frac{w_{\theta,\mathrm{bottom}}}{2}\int_{\Gamma_{\mathrm{bottom}}}
w_{\mathrm{rod}}(\mathbf{X})
\left(\theta^n-\tilde{\theta}_{\mathrm{bottom}}^n\right)^2\,d\Gamma
+
\frac{w_F}{2}
\left(F_x^n-\tilde{F}_x^n\right)^2
\Bigg] .
\end{aligned}
\end{equation}
Here, $\mathbf{u}_{\parallel}$ denotes the in-plane top-surface displacement components, $\tilde{\theta}_{\mathrm{bottom}}$ denotes the measured specimen temperature near the rod-contact region, and $F_x$ is the reaction force on the loaded boundary. The average displacement-curve term is set to zero in this case because the loaded-face displacement history at $x=L_0$ is already imposed as the mechanical boundary condition during the inverse problem.

For this synthetic inverse problem, the control variables are the ground-state shear modulus, the thermal expansion coefficient, and the thermal conductivity,
\begin{equation}
\mathbf{m}
=
\left[
G_{\mathrm{shear},0},
\alpha,
k_{\mathrm{therm}}
\right].
\end{equation}
The corresponding ground-truth values of the identified parameters are
\begin{equation}
G_{\mathrm{shear},0}=280~\mathrm{kPa},\qquad
\alpha=1.8\times 10^{-4}~\mathrm{K}^{-1},\qquad
k_{\mathrm{therm}}=400~\mathrm{mN/(s\,K)} .
\end{equation}
The specific heat, density, and contact coefficient are held fixed,
\begin{equation}
c_v=3600~\mathrm{J/(kg\,K)},\qquad
\rho = 10^{-6}~\mathrm{kg/mm^3},
\qquad
h_{\mathrm{contact}}=10000~\mathrm{mN/(mm\,s\,K)} .
\end{equation}
The inverse solve is initialized using the common perturbed starting point described above; explicitly,
\begin{equation}
\boldsymbol{m}^{(0)} = 1.1\,\boldsymbol{m}^{\mathrm{true}} .
\end{equation}
The rod/controller temperature history and the contact timing are treated as known prescribed inputs in this synthetic study. The corresponding adjoint gradients for the active inverse controls were verified against finite-difference gradients, as reported in Appendix~\ref{app:fd_syn2}.

\begin{figure}[H]
    \centering
    \begin{subfigure}[b]{0.49\linewidth}
        \centering
        \includegraphics[width=\linewidth]{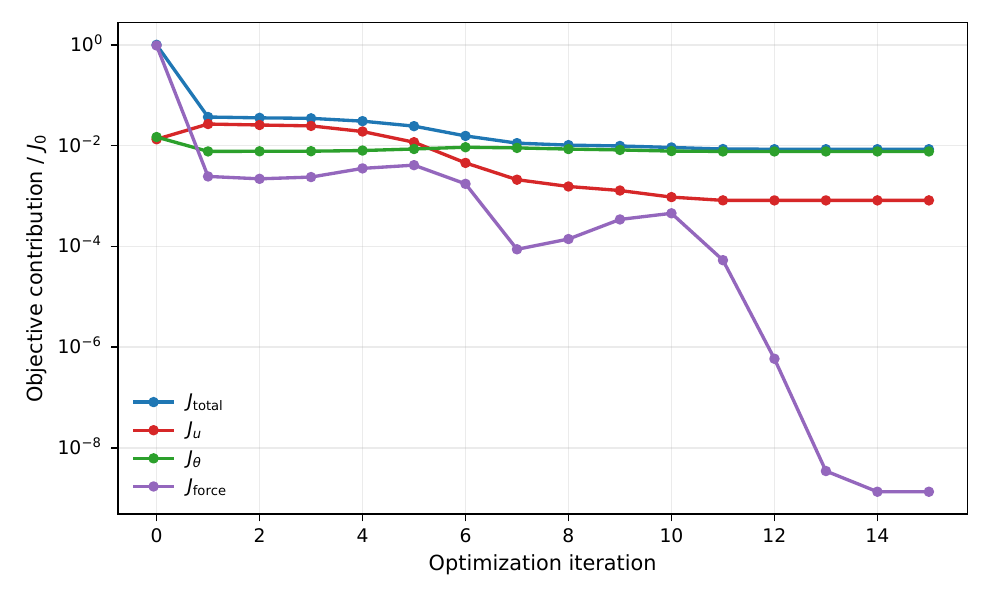}
        \caption{Objective convergence.}
    \end{subfigure}\hfill
    \begin{subfigure}[b]{0.49\linewidth}
        \centering
        \includegraphics[width=\linewidth]{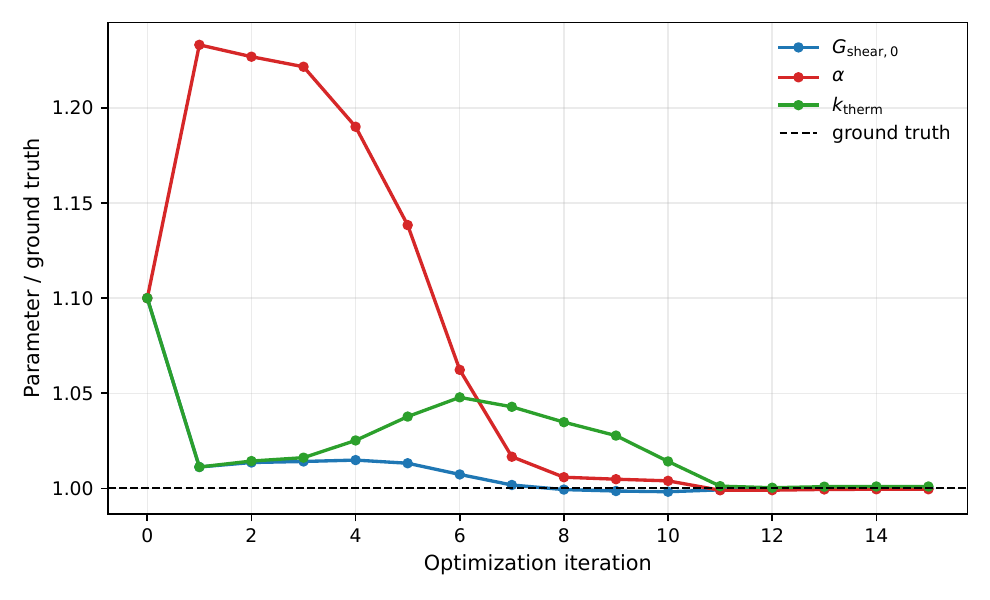}
        \caption{Parameter recovery.}
    \end{subfigure}

        \begin{subfigure}[b]{0.49\linewidth}
        \centering
        \includegraphics[width=\linewidth]{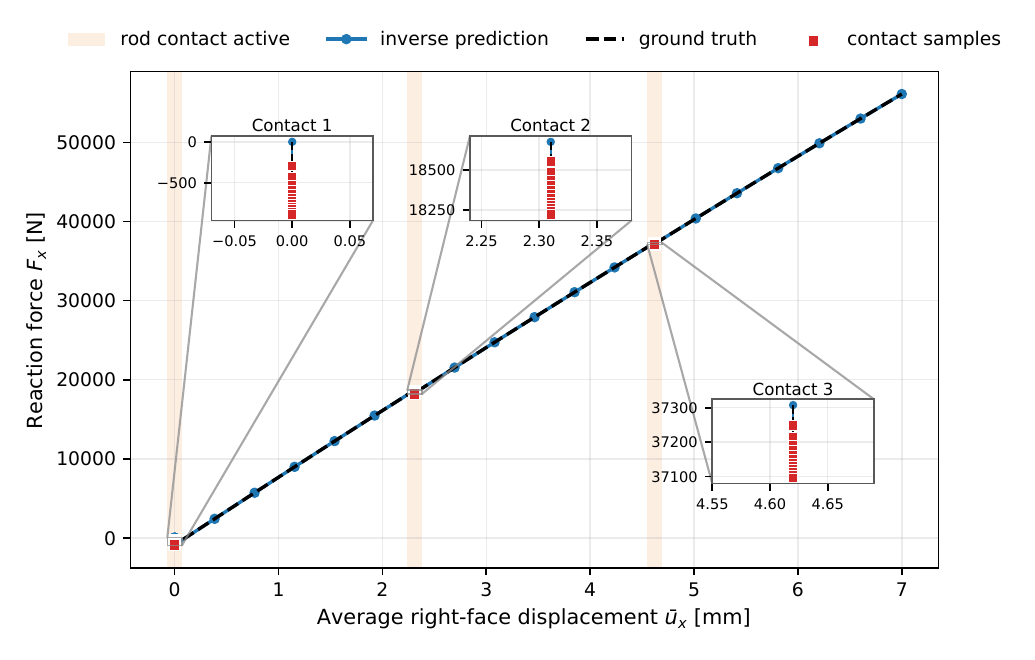}
        \caption{Force-displacement response. The inset axes magnify the intervals in which the rod is pushed down onto the specimen.}
    \end{subfigure}

    \caption{Inverse results for the staged rod-contact problem. The objective decreases during optimization, the identified shear modulus, thermal expansion coefficient, and thermal conductivity approach their ground-truth values, and the final force-displacement curve captures the perturbations caused by the repeated rod-contact events. In panel (c), the inset axes show the force-displacement response during the rod-contact intervals, when the rod is pushed down while the mechanical displacement is held fixed.}
    \label{fig:syn2_inverse_results}
\end{figure}

Figure~\ref{fig:syn2_inverse_results} shows that the inverse problem remains well behaved for the localized, transient heating protocol. Starting from the common perturbed initial guess, the normalized total objective decreases to approximately $8.5\times 10^{-3}$ of its initial value. The recovered shear modulus and thermal expansion coefficient converge very close to their ground-truth values, while the recovered thermal conductivity remains slightly above the truth at approximately $400.4$. The force-displacement response agrees closely with the synthetic reference curve and reproduces the local force perturbations associated with the rod-contact stages. The inset axes in Fig.~\ref{fig:syn2_inverse_results}(c) isolate the rod-contact intervals, where the rod is pushed down, and the displacement is held fixed, so that the thermally induced force changes can be seen more clearly. Compared with the uniform preconditioning problem, this case is more demanding because the localized heat input couples thermal transport, thermal expansion, and the mechanical reaction force over multiple loading stages.

\subsection{Experimental Data}
The experimental skin specimens were obtained from C57BL/6 mice. Mice were sacrificed in accordance with the NIH Guide for the Care and Use of Laboratory Animals under a University of Texas at Austin Institutional Animal Care and Use Committee (IACUC) approved protocol. Immediately following sacrifice, skin specimens were carefully excised from the dorsal (back) region and mounted in a custom thermo-mechanical test system. The system comprised a commercial biaxial tester (CellScale, Waterloo, Canada) retrofitted with a controllable temperature rod and two thermal cameras (FLIR A35 and A65, FLIR Systems, Wilsonville, OR, USA) \cite{meador2022biaxial}.

\begin{figure}
    \centering
    \includegraphics[width=0.5\linewidth]{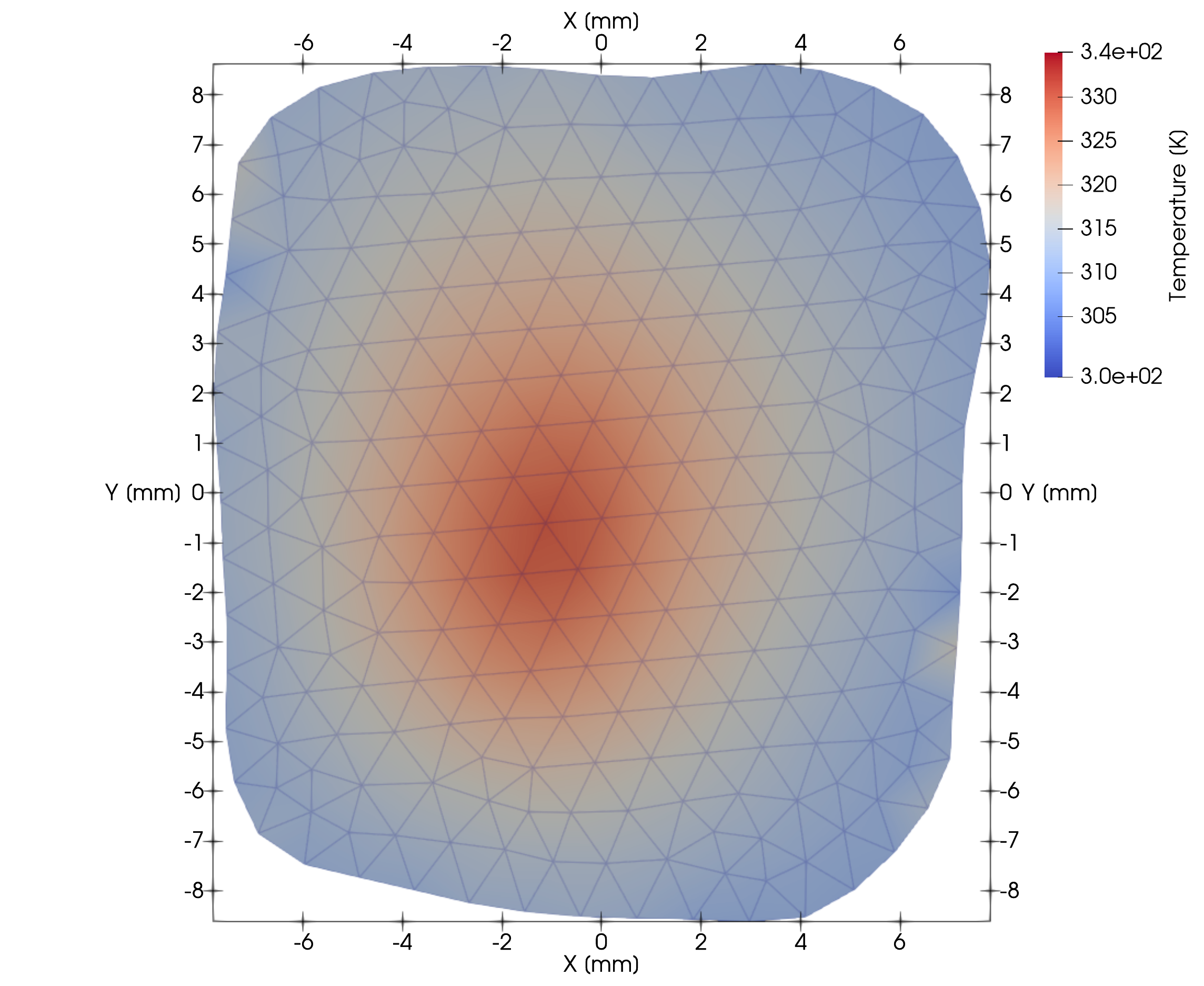}
    \caption{Dimensions of experimental specimen for the thermal test with superimposed measured temperatures at the top surface. The specimen is 353 $ \mu m$ thick.}
    \label{fig:expSpecimen}
\end{figure}

\begin{figure}[htbp]
    \centering
    \includegraphics[width=0.7\linewidth]{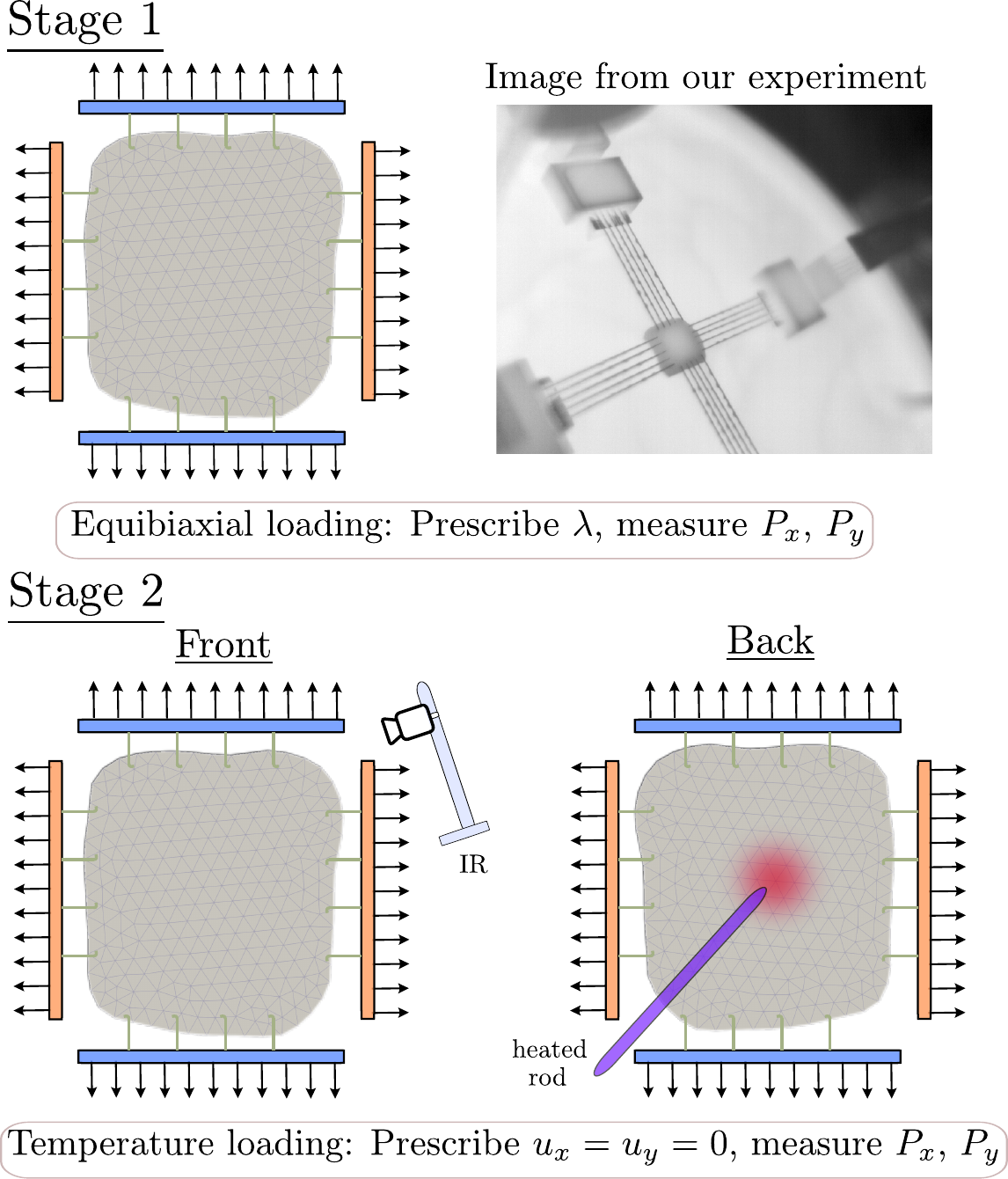}
    \caption{Experimental thermomechanical setup. The specimen is first characterized mechanically using equibiaxial loading--unloading data and is then subjected to thermal loading while the visible top-surface temperature field and in-plane boundary forces are recorded. The reported inverse problem does not include a thermal-stage DIC displacement field or a local temperature field around the rod-contact region.}
    \label{fig:experimental_setup}
\end{figure}

The experimental analysis is performed in two stages, see Figure \ref{fig:experimental_setup}. In the first stage, a mechanical baseline is obtained from cyclic equibiaxial loading data. The specimen is loaded and unloaded for ten cycles, and the tenth unloading branch is used for parameter identification. This choice reduces the influence of transient viscoelastic hysteresis and provides a stabilized quasi-equilibrium response for fitting the hyperelastic law. In the second stage, the calibrated mechanical response is held fixed and a thermal loading experiment is used to identify thermomechanical expansion and shrinkage parameters from the visible top-surface temperature field and boundary-force histories. The dimensions of the specimen are shown in Figure \ref{fig:expSpecimen}. The local temperature field around the rod-contact region is not included in the reported experimental inverse problem, and no DIC displacement field is used during the thermal stage. Accordingly, the experimental study is an incomplete-data specialization of the general full-field framework: the measured surface temperature is used for thermal data assimilation, while the boundary-force histories provide the primary mechanical constraints on the thermomechanical shrinkage response. The experimental setup and the measured data streams used in the inverse analysis are summarized in Fig.~\ref{fig:experimental_setup}.

\subsubsection{Experimental Constitutive Specialization}
The experimental model uses a single constitutive specialization for both stages of the analysis. The thermal-stage temperatures enter a range where skin response may involve thermally activated collagen denaturation and associated microstructural changes. Previous experiments on thermally denaturing skin showed anisotropic shrinkage, anisotropic force generation under constrained heating, and changes in mechanical response, with optical evidence linking these effects to collagen denaturation \cite{meador2022biaxial}. Corresponding mechanistic models have represented this behavior through denaturation kinetics, collagen-fiber coiling, viscoelastic relaxation, and damage-like loss of load-bearing capacity \cite{rausch2022biaxial}. The objective of the present experimental example is different: rather than resolving these biochemical and microstructural mechanisms explicitly, we use a reduced phenomenological thermomechanical specialization that captures the net temperature-induced expansion/shrinkage response observable through the measured surface temperature field and boundary-force histories. In this setting, the identified parameters should be interpreted as effective thermomechanical controls for the applied protocol, not as direct measurements of collagen denaturation kinetics.
Starting from the free-energy split in Eq.~\eqref{eq:free_energy_split}, the elastic response is represented by an isotropic specialization of the GOH-type exponential isochoric energy \cite{holzapfel2000new}, while thermal expansion and shrinkage are represented through a diagonal thermal distortion. Specifically, we define
\begin{equation}
\mathbf{F}_{\mathrm{th}}(\theta)
=
\mathrm{diag}
\left(
e^{\gamma_x(\theta)},
e^{\gamma_y(\theta)},
e^{\gamma_z(\theta)}
\right),
\qquad
\mathbf{F}_e=\mathbf{F}\mathbf{F}_{\mathrm{th}}^{-1},
\qquad
J_e=\frac{J}{J_{\mathrm{th}}},
\end{equation}
with
\begin{equation}
J_{\mathrm{th}}
=
\det \mathbf{F}_{\mathrm{th}}
=
\exp\left(\varepsilon_{\mathrm{th}}^v\right),
\qquad
\varepsilon_{\mathrm{th}}^v
=
\gamma_x+\gamma_y+\gamma_z .
\end{equation}
The isochoric invariant used in the exponential hyperelastic energy is then
\begin{equation}
\bar{I}_{1,e}
=
J_e^{-2/3}\mathrm{tr}(\mathbf{F}_e^{\top}\mathbf{F}_e).
\end{equation}
The experimental isochoric energy is
\begin{equation}
\Psi_{\mathrm{iso}}^{\mathrm{exp}}
=
\frac{\mu}{2}(\bar{I}_{1,e} - 3)
+
\frac{k_1}{2k_2}
\left[
\exp\left(k_2(\bar{I}_{1,e} - 3)^2\right) - 1
\right].
\end{equation}
Near-incompressibility in the coupled thermomechanical solve is enforced using the mixed pressure constraint in Eq.~\eqref{eq1:g}, with the thermal volumetric logarithmic strain $\varepsilon_{\mathrm{th}}^v$ defined above. Thus, the volumetric part of the coupled model uses the same thermal strain measure as the thermal distortion.

The directional thermal logarithmic strain components are chosen as
\begin{equation}
\begin{aligned}
\gamma_x(\theta) &= \alpha(\theta-\theta_0)-\beta_x s(\theta),\\
\gamma_y(\theta) &= \alpha(\theta-\theta_0)-\beta_y s(\theta),\\
\gamma_z(\theta) &= \alpha(\theta-\theta_0)-\frac{1}{2}(\beta_x+\beta_y)s(\theta),
\end{aligned}
\end{equation}
where the first term represents isotropic thermal expansion and the second term represents direction-dependent thermally activated shrinkage. The scalar activation function is taken as
\begin{equation}
s(\theta)
=
1-\exp\left(
-\frac{\langle \theta-\theta_{\mathrm{sh}}\rangle_+}{\Delta\theta_s}
\right),
\end{equation}
with a smooth positive-part approximation used in the numerical implementation. Thus $s(\theta)=0$ below the shrinkage activation temperature $\theta_{\mathrm{sh}}$ and approaches one at elevated temperature. This threshold is distinct from the actuator or rod setpoint temperature.
For the experimental inverse problem, the reference temperature is fixed at $\theta_0=297.55~\mathrm{K}$, the shrinkage activation temperature is fixed at $\theta_{\mathrm{sh}}=\theta_0+5~\mathrm{K}=302.55~\mathrm{K}$, and the saturation scale is fixed at $\Delta\theta_s=5~\mathrm{K}$. The smooth positive-part regularization uses a fixed smoothing scale of $1~\mathrm{K}$.

This specialization separates the calibrated mechanical response from the thermomechanical parameters identified in the thermal stage. The mechanical parameter vector is
\begin{equation}
\boldsymbol{\eta}_{\mathrm{mech}}
=
\left[
\mu,\ k_1,\ k_2
\right],
\end{equation}
where $\mu$ governs the small-strain shear response and $k_1$ and $k_2$ control the magnitude and rate of exponential strain stiffening. The reported thermal inverse controls are $\alpha$, $\beta_x$, and $\beta_y$. The experimental inverse solve is initialized at
$\alpha_0=4.5\times10^{-4}~\mathrm{K}^{-1}$,
$\beta_{x,0}=3.0\times10^{-2}$, and
$\beta_{y,0}=7.0\times10^{-2}$.
The remaining thermal transport and boundary coefficients are held fixed. The values
$c_\theta=3.6~\mathrm{N/(mm^2\,K)}$ and
$k_{\mathrm{therm}}=0.4~\mathrm{N/(s\,K)}$
are chosen from reported skin thermal properties \cite{hasgall2022itis,steensma2021sar}: using representative skin values
$\rho\approx1109~\mathrm{kg/m^3}$ and
$c_p\approx3390~\mathrm{J/(kg\,K)}$ gives
$\rho c_p\approx3.76\times10^6~\mathrm{J/(m^3\,K)}
=3.76~\mathrm{N/(mm^2\,K)}$, which we round to
$c_\theta=3.6~\mathrm{N/(mm^2\,K)}$, while reported skin conductivities near
$0.37~\mathrm{W/(m\,K)}$ motivate
$k_{\mathrm{therm}}=0.4~\mathrm{N/(s\,K)}
=0.4~\mathrm{W/(m\,K)}$.
The environmental exchange coefficient
$h_{\mathrm{conv}}=8.0\times10^{-3}~\mathrm{N/(mm\,s\,K)}$
corresponds to $8~\mathrm{W/(m^2\,K)}$ and is chosen as an effective dry surface-exchange coefficient of the order of natural convection plus linearized radiative exchange in air \cite{de1997convective}.
Finally, $h_{\mathrm{meas,top}}=10^3~\mathrm{N/(mm\,s\,K)}$
and $h_{\mathrm{meas,bottom}}=0$ are inverse-problem choices rather than literature material constants: the top coefficient weakly assimilates the measured visible top-surface temperature field, whereas the bottom coefficient is set to zero because no bottom or near-rod temperature field is included in the reported experimental inverse problem.
The roles of the experimental material parameters are summarized in Table~\ref{tab:experimental_parameters}.

\begin{table}[htbp]
    \centering
    \caption{Fixed and calibrated parameters in the experimental analysis. The mechanical parameters are fitted from the tenth unloading branch and then held fixed during the thermal inverse problem. The reported thermal inverse results focus on the thermomechanical expansion and shrinkage controls.}
    \label{tab:experimental_parameters}
    \small
    \setlength{\tabcolsep}{4pt}
    \begin{tabular}{p{0.20\linewidth}p{0.36\linewidth}p{0.34\linewidth}}
        \hline
        Parameter & Role & Status in analysis \\
        \hline
        $\mu$ & Small-strain shear response & Fixed: $0.029077~\mathrm{MPa}$ \\
        $k_1$ & Exponential stiffening magnitude & Fixed: $0.004622~\mathrm{MPa}$ \\
        $k_2$ & Exponential stiffening rate & Fixed: $7.840417$ (dimensionless) \\
        $\alpha$ & Isotropic thermal expansion & Initial/prior: $4.5\times10^{-4}~\mathrm{K}^{-1}$; calibrated \\
        $\beta_x$ & Directional thermal shrinkage in $x$ & Initial/prior: $3.0\times10^{-2}$ (dimensionless); calibrated \\
        $\beta_y$ & Directional thermal shrinkage in $y$ & Initial/prior: $7.0\times10^{-2}$ (dimensionless); calibrated \\
        \hline
    \end{tabular}
\end{table}

\subsubsection{Mechanical Preconditioning and Hyperelastic Calibration}

During the mechanical calibration, the specimen is modeled isothermally at the reference temperature. In this limit $s(\theta_0)=0$, $\mathbf{F}_{\mathrm{th}}=\mathbf{I}$, and the experimental model reduces to the hyperelastic law above with no active thermal strain. In the full finite-element mechanical preconditioning problem, near-incompressibility is enforced by the volumetric penalty
\begin{equation}
\Psi_{\mathrm{vol}}^{\mathrm{mech}} = \frac{\kappa}{2}(J-1)^2,
\end{equation}
where $\kappa$ is the mechanical-only bulk penalty and is not fitted. It plays the same stabilizing role as $K$ in the coupled mixed formulation, but is used only in the uncoupled mechanical preconditioning solve. The reduced equibiaxial fit used to estimate $\boldsymbol{\eta}_{\mathrm{mech}}$ uses the corresponding incompressible specialization.

\begin{figure}[htbp]
    \centering
    \includegraphics[width=0.6\linewidth]{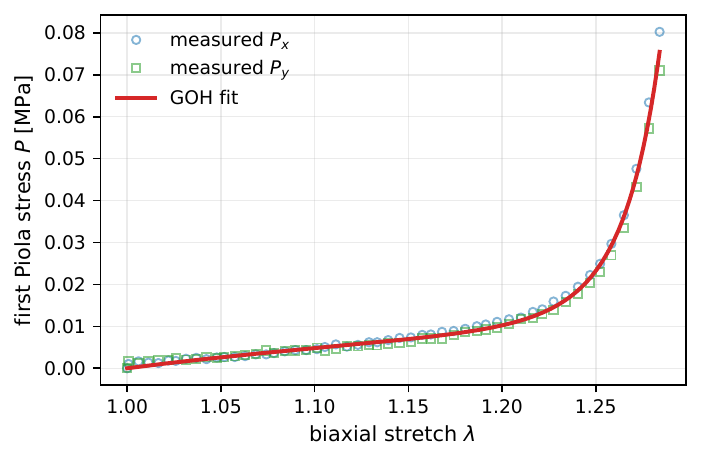}
    \caption{Mechanical calibration from the tenth unloading branch of the cyclic equibiaxial experiment. The fitted GOH response is compared with the measured nominal stresses $P_x$ and $P_y$.}
    \label{fig:experimental_goh_fit}
\end{figure}

For the equibiaxial calibration, the in-plane stretches are approximated as
\begin{equation}
\lambda_1=\lambda_2=\lambda,
\qquad
\lambda_3=\lambda^{-2},
\end{equation}
so that
\begin{equation}
I_1 = 2\lambda^2 + \lambda^{-4}.
\end{equation}
Let
\begin{equation}
\left\{
(\lambda^{(n)},P_x^{(n)},P_y^{(n)})
\right\}_{n=1}^{N}
\end{equation}
denote the measured stretch and nominal in-plane stresses on the tenth unloading branch. These calibration quantities are stresses and are distinct from the thermal-stage boundary force resultants reported below. The isotropic model predicts
\begin{equation}
P_x^{\mathrm{model}}(\lambda;\boldsymbol{\eta}_{\mathrm{mech}})
=
P_y^{\mathrm{model}}(\lambda;\boldsymbol{\eta}_{\mathrm{mech}})
=
P_{\mathrm{biax}}(\lambda;\boldsymbol{\eta}_{\mathrm{mech}}),
\end{equation}
and the mechanical parameters are obtained from
\begin{equation}
\boldsymbol{\eta}_{\mathrm{mech}}^{*}
=
\operatorname*{arg\,min}_{\boldsymbol{\eta}_{\mathrm{mech}}}
\frac{1}{N}
\sum_{n=1}^{N}
\left[
\left(P_{\mathrm{biax}}(\lambda^{(n)};\boldsymbol{\eta}_{\mathrm{mech}})-P_x^{(n)}\right)^2
+
\left(P_{\mathrm{biax}}(\lambda^{(n)};\boldsymbol{\eta}_{\mathrm{mech}})-P_y^{(n)}\right)^2
\right].
\end{equation}
The fitted values used in the thermomechanical inverse problem are
\begin{equation}
\mu = 0.029077~\mathrm{MPa},
\qquad
k_1 = 0.004622~\mathrm{MPa},
\qquad
k_2 = 7.840417\quad\text{(dimensionless)} .
\end{equation}
These mechanical parameters are held fixed during the subsequent thermal inverse analysis.
The resulting mechanical fit is shown in Fig.~\ref{fig:experimental_goh_fit}.

\subsubsection{Thermal Loading and Data Reduction}

After mechanical calibration, the specimen is subjected to thermal loading while the in-plane boundary forces and the visible top-surface temperature field are recorded \cite{pitarresi2003review,chrysochoos2012infrared}. The thermal measurements are exported onto the corresponding experimental finite element surface mesh before the inverse solve. Specifically, the thermography data are stored as nodal temperature fields on the reconstructed surface mesh at each recorded frame; the inverse solver reads the visible top-surface node set and uses those nodal values as $\tilde{\theta}_{\mathrm{top}}^n$. If the thermal image frames and force-recording samples are not already indexed identically, the temperature frames are aligned to the force history using the recorded frame indices or the nearest available time. Invalid or unavailable temperature values are excluded through the mask $\chi_{\mathrm{top}}^n$. The measured force histories are zero-referenced to the first thermal-frame sample, and both in-plane force components are retained for the inverse problem. The thermal-stage data used here do not include a corresponding DIC displacement field, and the local temperature field around the rod-contact region is not included in the inverse analysis. Consequently, displacement-field, displacement-curve, and near-rod temperature misfit terms are disabled in this experimental inverse problem.

\begin{figure}[htbp]
    \centering
    \includegraphics[width=0.7\linewidth]{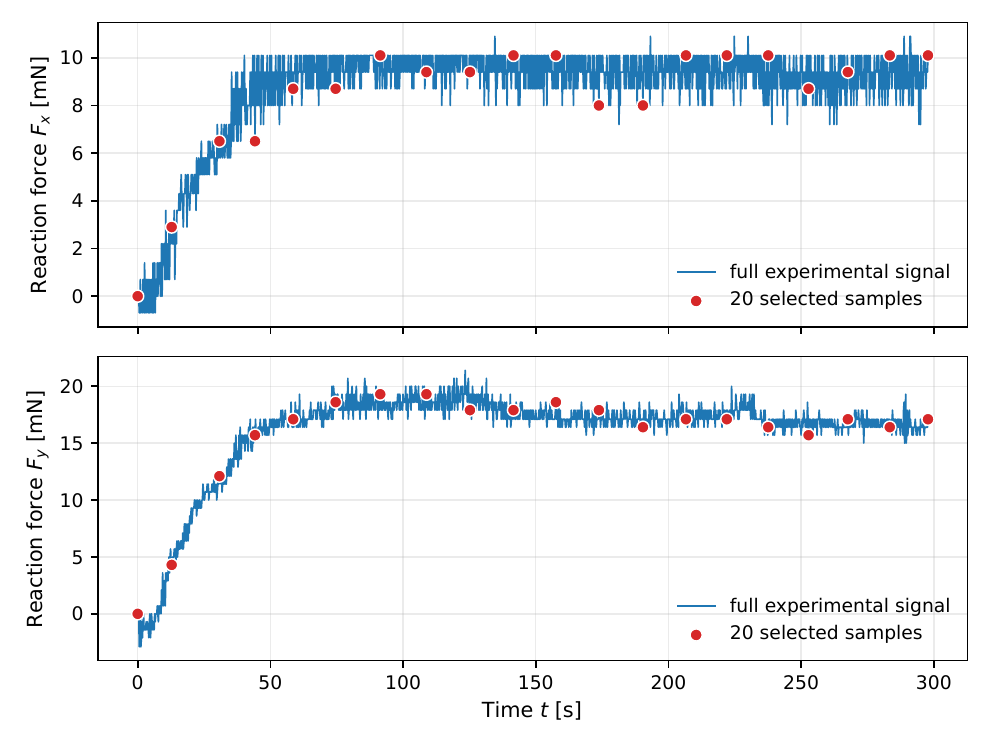}
    \caption{Experimental force-history downselection for the thermal inverse problem. The continuous curves show the measured zero-referenced in-plane force resultants $P_x$ and $P_y$, and the markers indicate the selected time points used in the objective.}
    \label{fig:experimental_force_downselection}
\end{figure}

The full thermal experiment contains more recorded time steps than are needed for the inverse solve. We therefore downselect the objective evaluation points from the measured force history, retaining representative samples across the thermal loading history. Figure~\ref{fig:experimental_force_downselection} shows the measured in-plane force-resultant histories $P_x$ and $P_y$ together with the selected time points used in the inverse objective.

\subsubsection{Thermomechanical Inverse Problem}

The thermal inverse problem uses the experimental constitutive specialization above with the calibrated GOH parameters held fixed. The measured force histories in Fig.~\ref{fig:experimental_force_downselection} show different responses in the two in-plane directions during thermal loading \cite{meador2022biaxial,rausch2022biaxial}, motivating the anisotropic shrinkage amplitudes $\beta_x$ and $\beta_y$ rather than a single isotropic shrinkage parameter. The inverse problem, therefore, identifies the thermomechanical parameters $\alpha$, $\beta_x$, and $\beta_y$ from the thermal-stage temperature and force data, while the mechanical parameters $\mu$, $k_1$, and $k_2$ remain fixed at the values obtained from the tenth unloading branch.

\begin{figure}[H]
    \centering
    \begin{subfigure}[b]{0.32\linewidth}
        \centering
        \includegraphics[width=\linewidth]{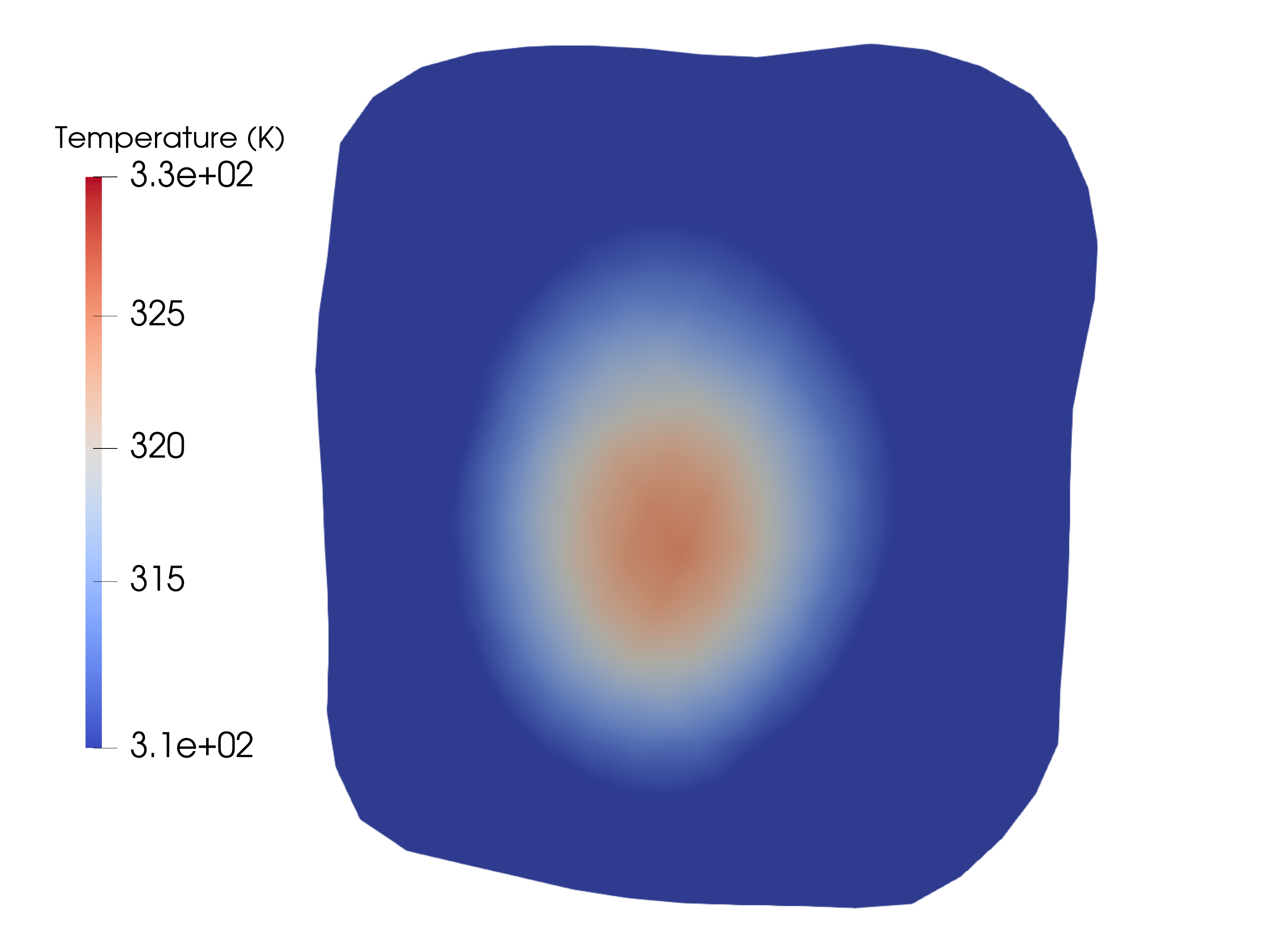}
        \caption{$t=13\, \text{s}$}
    \end{subfigure}\hfill
    \begin{subfigure}[b]{0.32\linewidth}
        \centering
        \includegraphics[width=\linewidth]{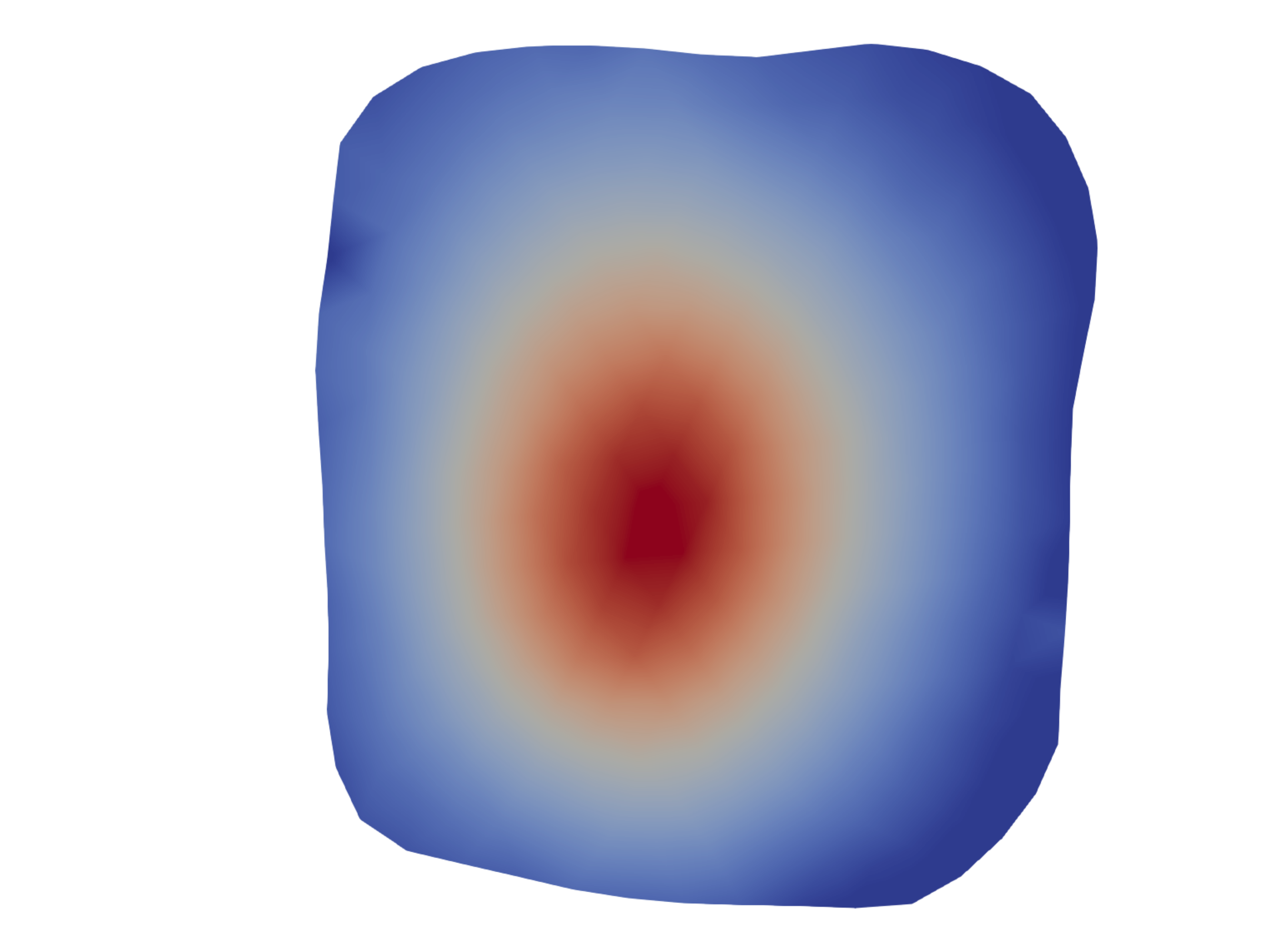}
        \caption{$t=74.6\, \text{s}$}
    \end{subfigure}\hfill
    \begin{subfigure}[b]{0.32\linewidth}
        \centering
        \includegraphics[width=\linewidth]{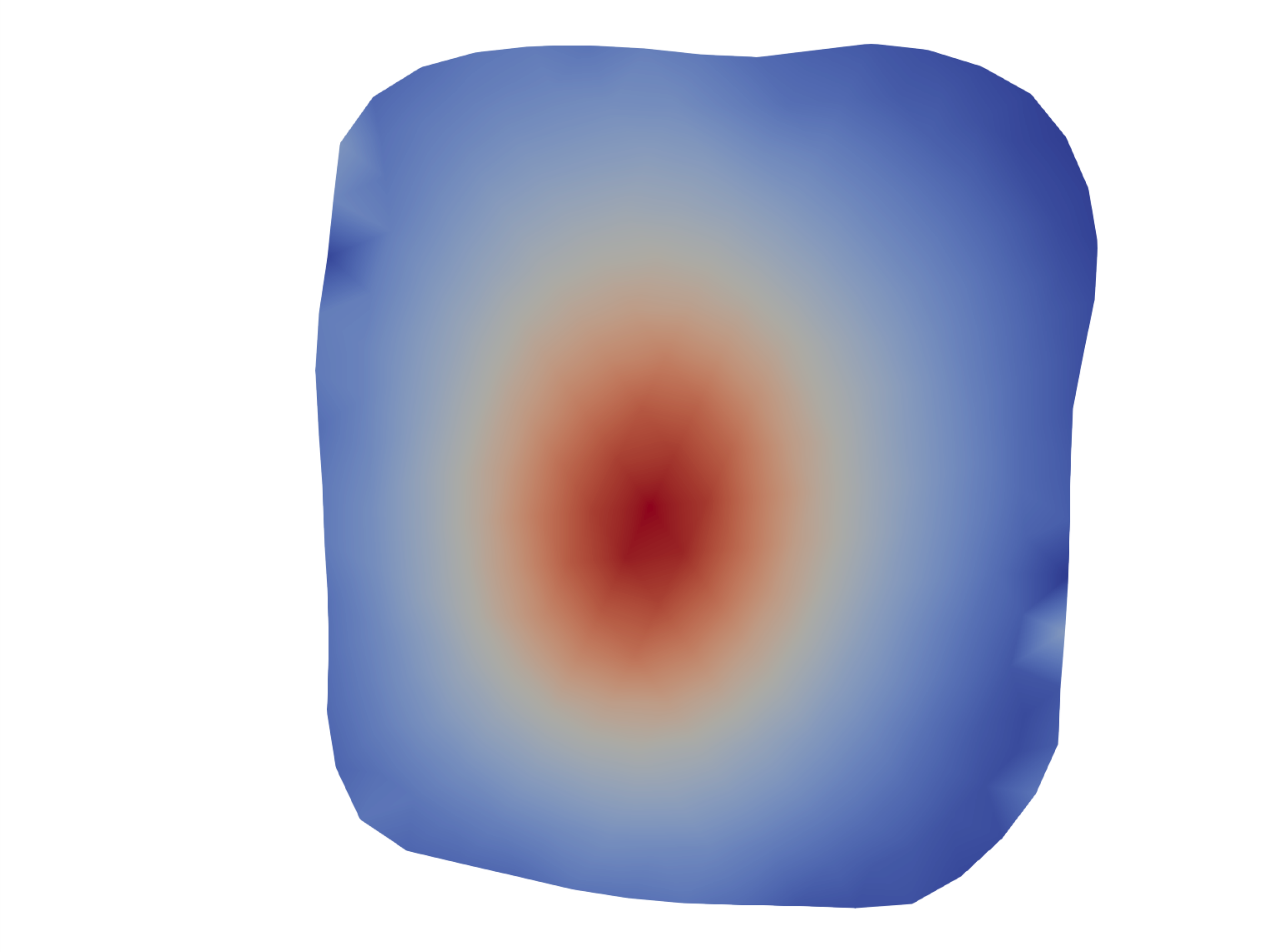}
        \caption{$t=254.4\, \text{s}$}
    \end{subfigure}

    \begin{subfigure}[b]{0.32\linewidth}
        \centering
        \includegraphics[width=\linewidth]{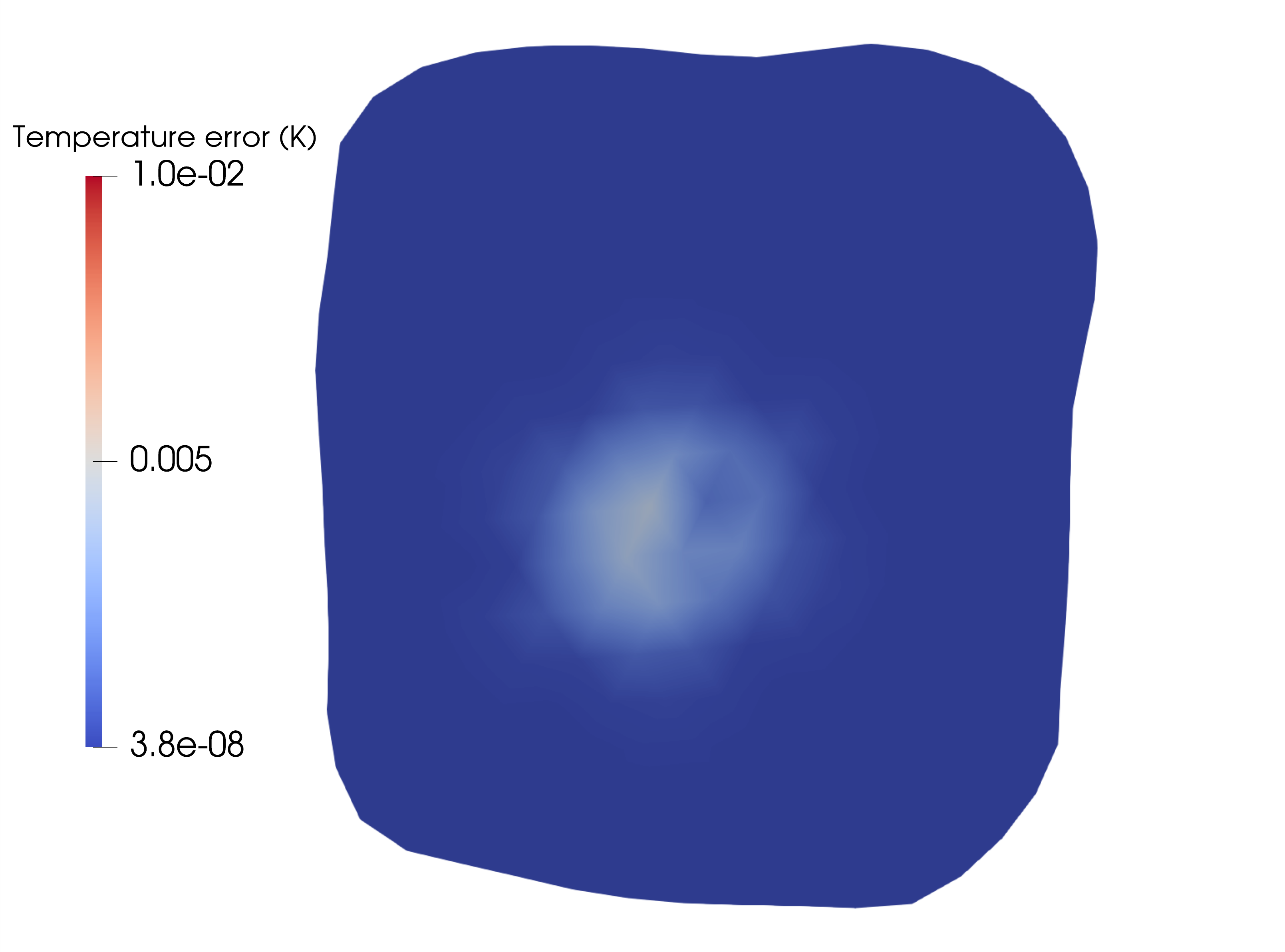}
        \caption{$t=13\, \text{s}$}
    \end{subfigure}\hfill
    \begin{subfigure}[b]{0.32\linewidth}
        \centering
        \includegraphics[width=\linewidth]{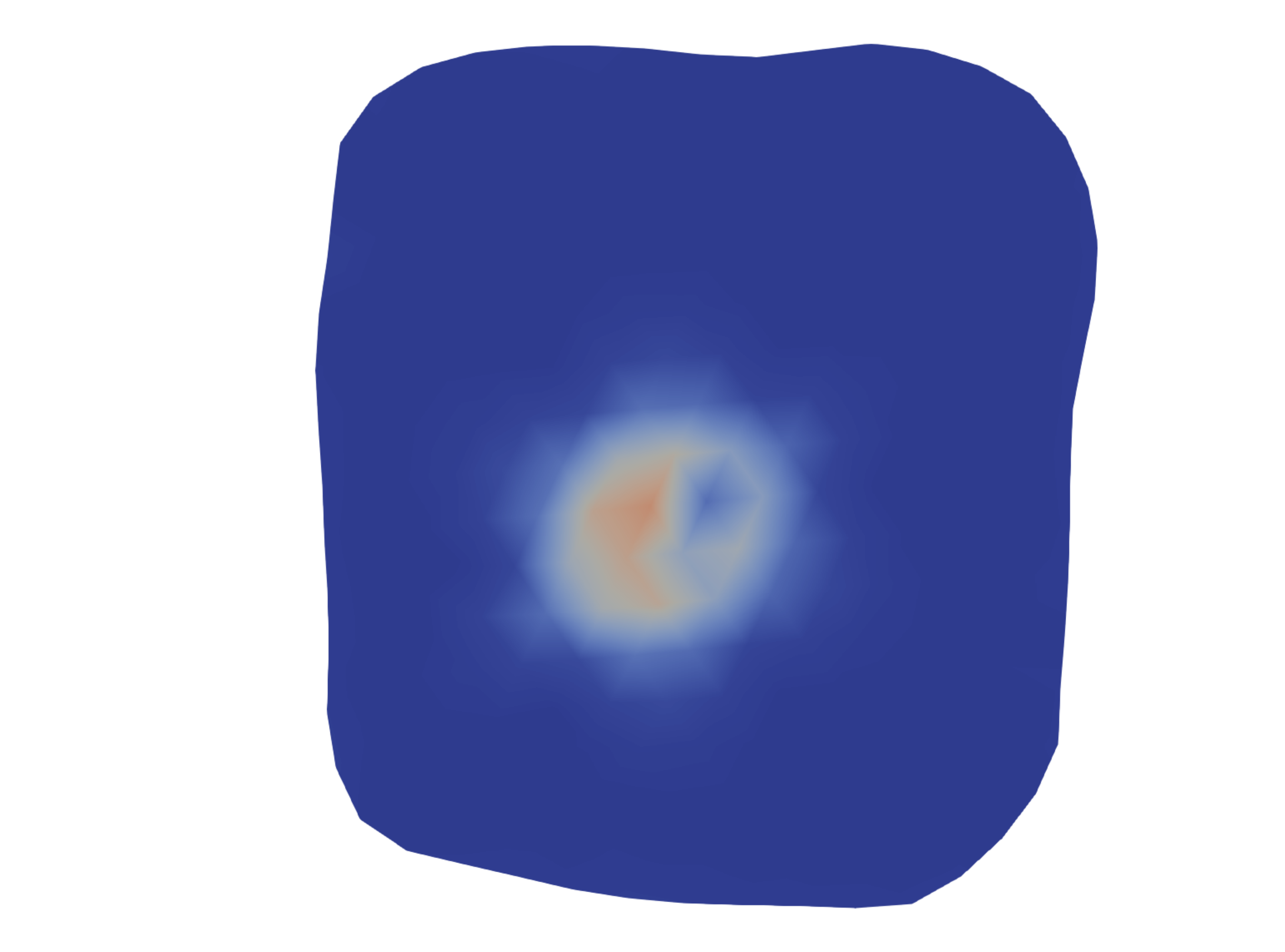}
        \caption{$t=74.6\, \text{s}$}
    \end{subfigure}\hfill
    \begin{subfigure}[b]{0.32\linewidth}
        \centering
        \includegraphics[width=\linewidth]{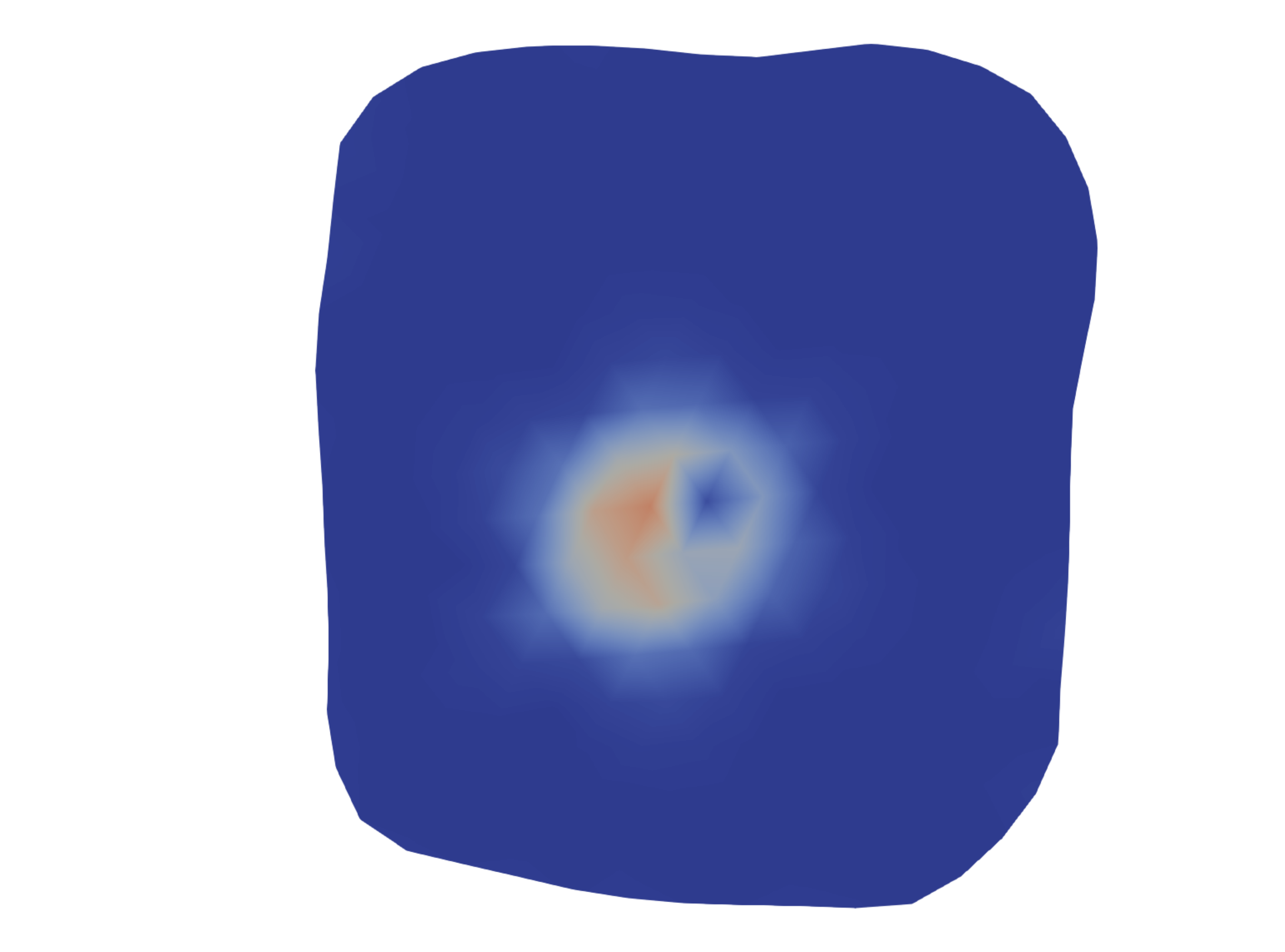}
        \caption{$t=254.4\, \text{s}$}
    \end{subfigure}

    \caption{IR-measured top-surface temperature fields projected onto the computational mesh at selected time points (a)--(c) and temperature-assimilation residuals (d)--(f) for the experimental thermal inverse problem. Panels (a)--(c) correspond to $t=13.0~\mathrm{s}$, $t=74.6~\mathrm{s}$, and $t=254.4~\mathrm{s}$, respectively. Panels (d)--(f) show the corresponding absolute temperature differences, $|\theta-\tilde{\theta}_{\mathrm{top}}|$, between the calibrated inverse solution and the projected measured temperature fields.}
    \label{fig:truevsPred}
\end{figure}

Figure~\ref{fig:truevsPred} shows representative top-surface temperature fields used in the experimental thermal inverse problem. The projected IR measurements exhibit a localized heated region that evolves over the selected time points. These measured temperature fields enter the inverse formulation both through the Robin temperature-assimilation boundary condition and through the top-surface temperature mismatch term in the objective. The corresponding residual maps show that the calibrated inverse solution remains closely aligned with the measured top-surface temperature field on the observed surface. Because this same temperature field is used to guide the thermal solve, these residuals should be interpreted as a consistency check of the temperature assimilation and objective matching, rather than as an independent validation of full-field thermal prediction.

The experimental thermal problem is data-driven but restricted to the measured top-surface thermal field. Because the local rod-contact temperature field is not included, the reported inverse problem does not prescribe or fit a bottom near-rod temperature field. Instead, the measured top-surface temperature data are incorporated through a Robin-type measurement assimilation term,
\begin{equation}
-k_{\mathrm{therm}}\nabla\theta\cdot\mathbf{N}
=
h_{\mathrm{meas}}\,\chi_{\mathrm{top}}^n
\left(\theta-\tilde{\theta}_{\mathrm{top}}^n\right)
\quad \text{on } \Gamma_{\mathrm{top}},
\end{equation}
where $\chi_{\mathrm{top}}^n$ masks the available top-surface temperature data at time step $n$.
The coefficient $h_{\mathrm{meas}}$ is a fixed assimilation penalty rather than a material parameter or an identified control. In the reported experimental calculations we use
$h_{\mathrm{meas}}=h_{\mathrm{meas,top}}=10^3~\mathrm{N/(mm\,s\,K)}$,
which can be interpreted through the associated thermal penalty length
\begin{equation}
\ell_{\mathrm{meas}}
=
\frac{k_{\mathrm{therm}}}{h_{\mathrm{meas,top}}}
=
4.0\times10^{-4}~\mathrm{mm}.
\end{equation}
This length scale is much smaller than both the specimen thickness and the surface mesh spacing, so the measured top-surface temperature is enforced strongly wherever valid thermography data are available.
This weak Robin treatment is used instead of a strong Dirichlet prescription because experimental infrared temperature fields can be noisy, spatially incomplete, and affected by registration uncertainty. The assimilation term, therefore, allows the thermal solve to be guided by the measured surface temperature while still regularizing the field through the heat equation and the available boundary masks. The same measured top-surface temperature field is also compared in the objective to control the temperature-field mismatch. This retained temperature-misfit term represents the more general case in which the Robin assimilation coefficient is finite and the measured field is not fully enforced; for the large value of $h_{\mathrm{meas}}$ used here, it mainly serves as a consistency term, while the force histories provide the primary calibration information for the thermomechanical shrinkage parameters. Consequently, the experimental temperature response should be interpreted as boundary-supported thermal data assimilation rather than a blind prediction of the full temperature field.

The active experimental inverse objective is
\begin{equation}
\begin{aligned}
J_{\mathrm{exp}}
=
\sum_{n=1}^{N}
\Bigg[
\frac{w_\theta}{2}
\int_{\Gamma_{\mathrm{top}}}
\chi_{\mathrm{top}}^n
\left(\theta^n-\tilde{\theta}_{\mathrm{top}}^n\right)^2\,d\Gamma
+
\frac{w_F}{2}
\left[
\left(P_x^n-\tilde{P}_x^n\right)^2
+
\left(P_y^n-\tilde{P}_y^n\right)^2
\right]
\Bigg].
\end{aligned}
\end{equation}
Thus, the temperature term penalizes the residual in the predicted visible top-surface thermal field, while the two force components constrain the thermomechanical shrinkage response. No separate bottom or near-rod temperature objective term is used in the reported experimental inverse problem. Because the rod-contact temperature field is not available, the bottom face is not prescribed by a rod-temperature boundary condition in the inverse solve; with $h_{\mathrm{meas,bottom}}=0$, it is treated as an unobserved natural thermal boundary, while the measured top-surface thermography enters through the Robin assimilation term above. The adjoint gradients for the three active experimental controls in this objective are verified against central finite differences in Appendix~\ref{app:fd_exp}.

\subsubsection{Experimental Inverse Results}

\begin{figure}[htbp]
    \centering
    \begin{subfigure}[b]{0.49\linewidth}
        \centering
        \includegraphics[width=\linewidth]{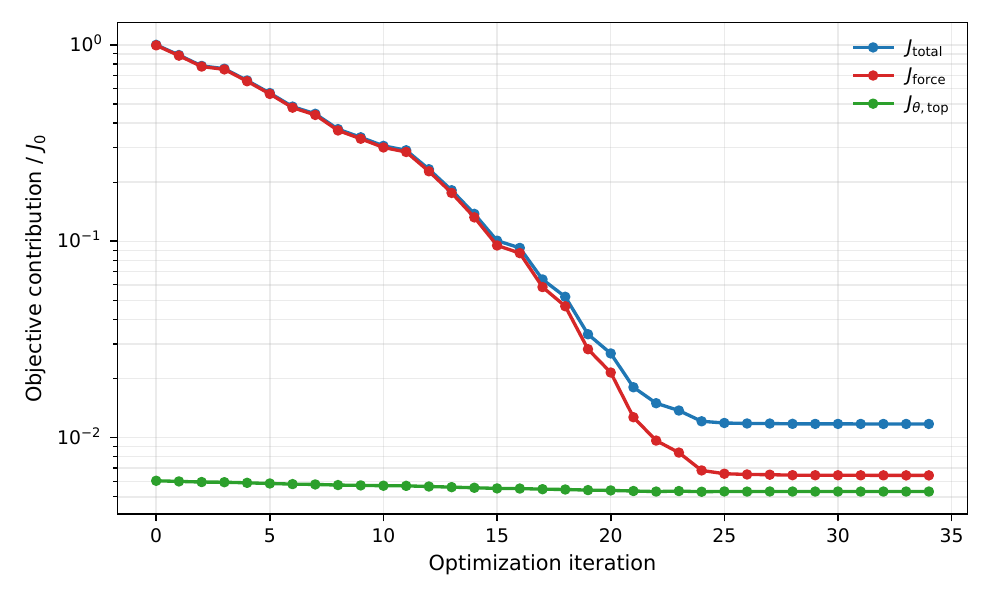}
        \caption{Objective convergence.}
    \end{subfigure}\hfill
    \begin{subfigure}[b]{0.49\linewidth}
        \centering
        \includegraphics[width=\linewidth]{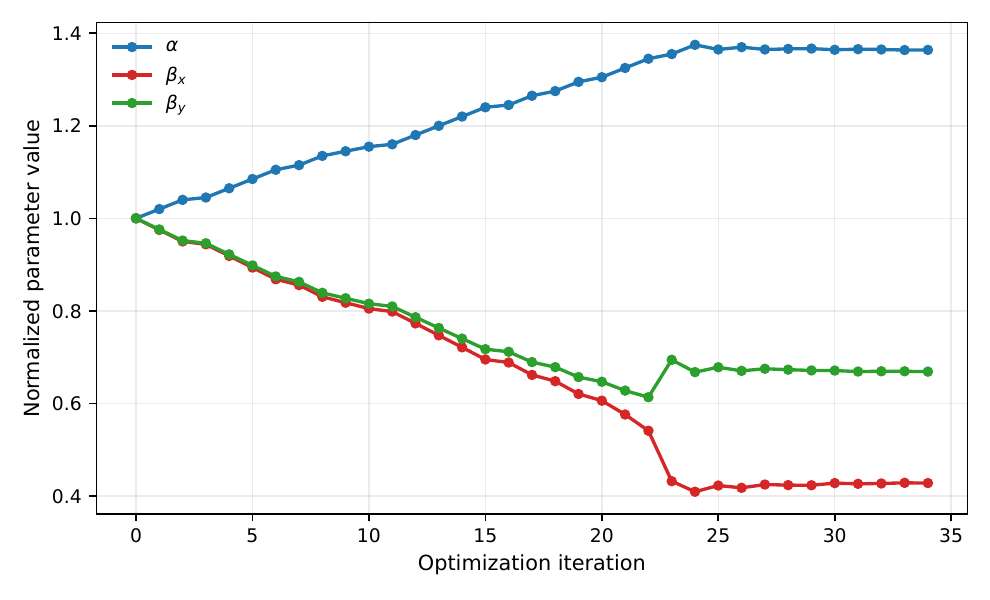}
        \caption{Thermomechanical controls.}
    \end{subfigure}

    \begin{subfigure}[b]{0.49\linewidth}
        \centering
        \includegraphics[width=\linewidth]{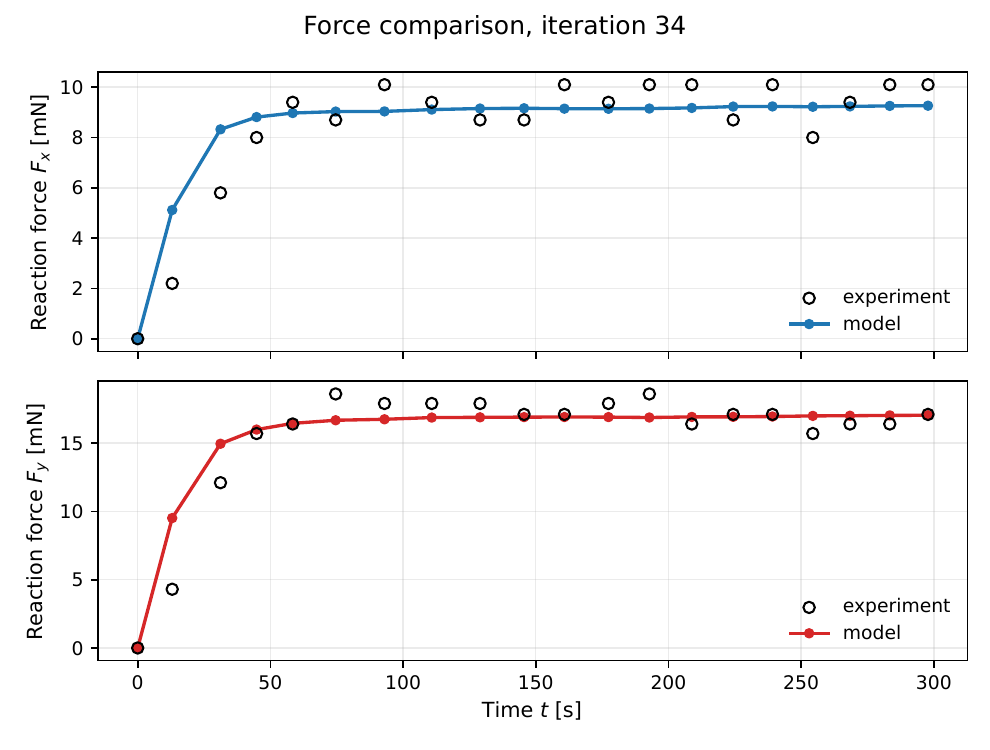}
        \caption{Force-history comparison.}
    \end{subfigure}

    \caption{Experimental thermomechanical inverse results. The objective decreases over the inverse iterations, the reported thermomechanical controls $\alpha$, $\beta_x$, and $\beta_y$ evolve from their initial values, and the final model prediction is compared with the measured in-plane force histories. The reported objective uses the measured top-surface temperature field and force histories, without a near-rod temperature objective.}
    \label{fig:experimental_inverse_results}
\end{figure}

Figure~\ref{fig:experimental_inverse_results} summarizes the experimental inverse fit. The normalized total objective decreases to approximately $1.2\times 10^{-2}$ of its initial value, with most of the reduction coming from the force-history term. The identified thermal expansion coefficient increases from its prior value, while the directional shrinkage amplitudes decrease and settle to different values in the two in-plane directions. This result is consistent with the anisotropic force response visible in Fig.~\ref{fig:experimental_force_downselection}, where the measured $P_x$ and $P_y$ histories differ during thermal loading. The final force comparison captures the dominant evolution of both $P_x$ and $P_y$, demonstrating that the calibrated hyperelastic baseline combined with the identified thermomechanical shrinkage parameters can reproduce the main experimental force response during thermal loading.

\section{Conclusions and Outlook}\label{sec:conclusions}

We presented a finite-strain PDE-constrained inverse framework for identifying thermomechanical material parameters from surface displacement, surface temperature, and boundary-force measurements. The forward model couples a near-incompressible thermo-hyperelastic formulation with heat conduction and thermoelastic source terms derived from a common free energy. The inverse problem uses adjoint sensitivities and scaled gradient-based optimization to combine heterogeneous measurement data in a single objective.
The synthetic studies verify the internal consistency of the inverse formulation under controlled conditions. A uniform thermal-preconditioning problem demonstrates recovery of shear modulus and thermal expansion parameters from mechanical observations, while a staged rod-contact problem demonstrates recovery under localized transient heating and repeated mechanical holds. The experimental application then shows how the same framework can be applied to measured thermomechanical data by first calibrating a stabilized hyperelastic response from cyclic equibiaxial loading and then identifying thermal expansion and directional shrinkage parameters from thermal-stage force and temperature histories.
The current approach has several limitations. The synthetic examples do not include added measurement noise or model-form error, so they test consistency and recoverability rather than robustness under all experimental uncertainties. The mechanical response is treated as hyperelastic after preconditioning, so rate-dependent viscoelastic effects are not modeled explicitly. Furthermore, some thermal quantities, such as heat-transfer coefficients, contact conditions, specific heat, and thermal conductivity, can be difficult to identify simultaneously from the current data streams. In the experimental application, the thermal shrinkage law is an effective phenomenological representation of the measured response; it does not resolve collagen denaturation kinetics, biochemical state evolution, viscoelastic relaxation, or damage mechanisms explicitly.

Future work should address these limitations by incorporating measurement-noise models, uncertainty quantification, and stronger regularization for partially identifiable parameters. Extending the constitutive model to include internal variables for denaturation, viscoelasticity, irreversible shrinkage, or damage would allow the mechanical calibration and thermal inverse problem to account for rate-dependent and microstructurally motivated thermal effects directly. Beyond adding individual mechanisms to the present parametric law, the constitutive ansatz could also be made more flexible by replacing selected parts of the prescribed free energy with constrained data-driven representations, such as polyconvex thermoelastic free-energy models that preserve thermodynamic structure while learning richer deformation- and temperature-dependent responses \cite{fuhg2024polyconvex}. A more detailed actuator/contact model, possibly including identification of effective heat-transfer parameters, would reduce reliance on surface-temperature assimilation to represent unobserved contact heating. Additional experimental data, such as synchronized DIC during thermal loading, richer thermal boundary measurements, or repeated tests under different heating protocols, would further improve identifiability. A natural next step is to apply the inverse model across multiple specimens and estimate population-level thermomechanical parameters, either by averaging specimen-wise inverse results or by solving a joint inverse problem with specimen-specific nuisance parameters and shared population-level controls. These extensions would move the framework toward robust calibration of thermomechanical material models from realistic, noisy, incomplete, and specimen-variable experimental observations.

\section*{Acknowledgements}
This material is based upon work partially supported by the U.S. National Science Foundation under award No. 2452029 (JNF and JY).  
The opinions, findings, and conclusions, or recommendations expressed are those of the authors and do not necessarily reflect the views of the NSF.
\\
The authors acknowledge the Texas Advanced Computing Center (TACC) at The University of Texas at Austin for providing computational resources that have contributed to the research results reported within this paper.
\\
This work was supported by the Laboratory Directed Research and Development program at Sandia National Laboratories, a multimission laboratory managed and operated by National Technology and Engineering Solutions of Sandia LLC, a wholly owned subsidiary of Honeywell International Inc. for the U.S. Department of Energy’s National Nuclear Security Administration under contract DE-NA0003525. This written work is authored by an employee of NTESS. The employee, not NTESS, owns the right, title and interest in and to the written work and is responsible for its contents. Any subjective views or opinions that might be expressed in the written work do not necessarily represent the views of the U.S. Government. The publisher acknowledges that the U.S. Government retains a non-exclusive, paid-up, irrevocable, world-wide license to publish or reproduce the published form of this written work or allow others to do so, for U.S. Government purposes. The DOE will provide public access to results of federally sponsored research in accordance with the DOE Public Access Plan. 

\bibliographystyle{unsrt}
\bibliography{bib}

\appendix
\section{Verification of Adjoint Gradients}\label{app:fd_checks}
The gradients used in the reduced-space optimization are checked by comparing the adjoint derivative of the reduced objective with finite-difference derivatives. For each selected control $m_i$, the adjoint calculation gives $g_{\mathrm{AD}}=\partial J/\partial m_i$ at the initial inverse iterate, with all other controls fixed. The finite-difference derivative is then computed from fresh objective evaluations using a central stencil,
\begin{equation}
g_{\mathrm{FD}}(\epsilon)
=
\frac{J(m_i+\delta_i)-J(m_i-\delta_i)}{2\delta_i},
\qquad
\delta_i=\epsilon s_i ,
\end{equation}
where $\epsilon$ is the nondimensional finite-difference sweep parameter and $s_i$ is the control-specific perturbation scale. Small-magnitude controls, such as $\alpha$, are perturbed relative to their current or prior magnitude, while controls of order one or larger use an absolute perturbation scale. The plots below report both the absolute gradient error $|g_{\mathrm{FD}}-g_{\mathrm{AD}}|$ and the relative error. The expected behavior is a minimum over an intermediate range of $\epsilon$, with roundoff and nonlinear-solver noise dominating for very small perturbations and truncation error dominating for large perturbations.

\clearpage

\subsection{Synthetic Problem 1}\label{app:fd_syn1}

\begin{figure}[htbp]
    \centering
    \begin{subfigure}[b]{0.49\linewidth}
        \centering
        \includegraphics[width=\linewidth]{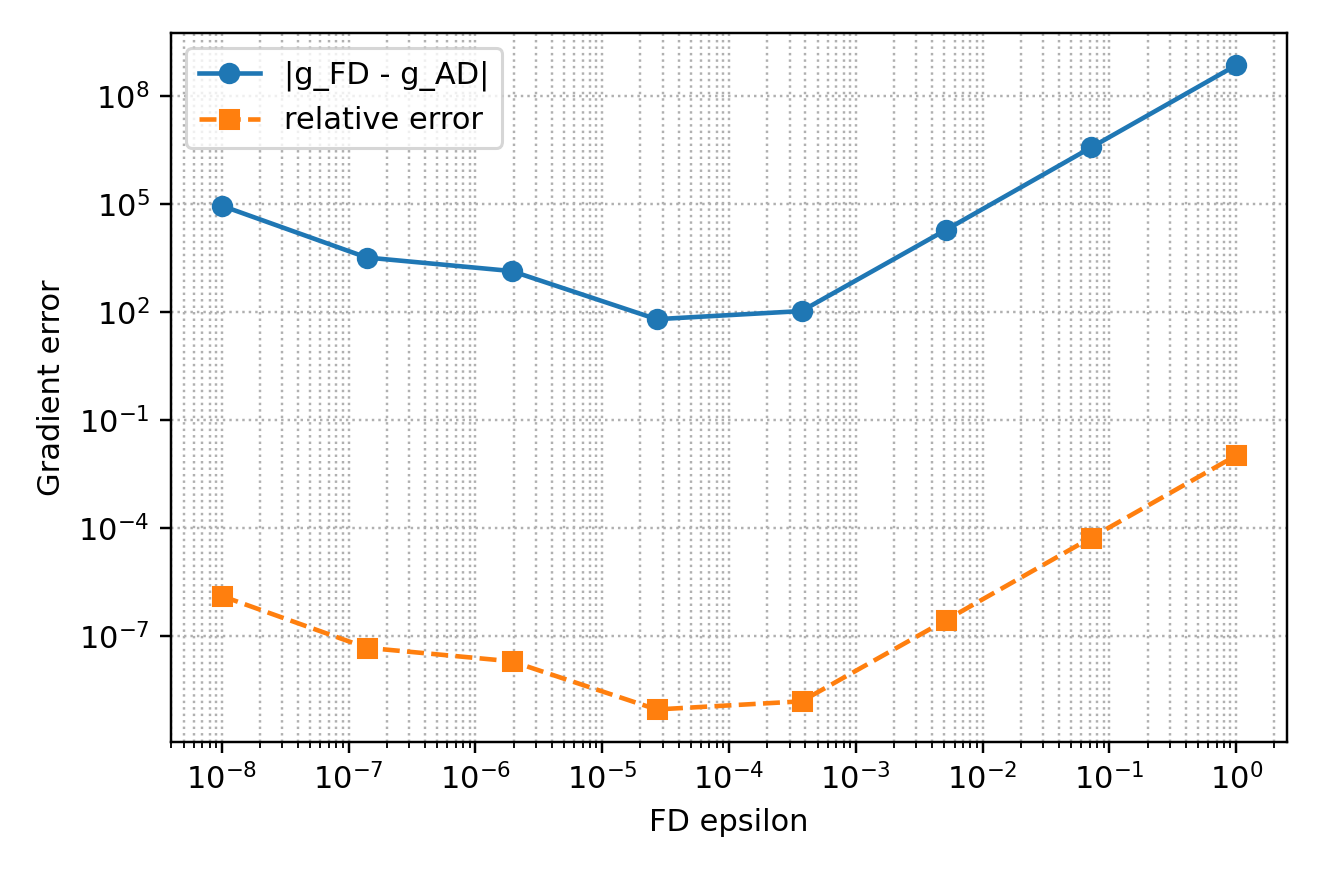}
        \caption{$\alpha$}
    \end{subfigure}\hfill
    \begin{subfigure}[b]{0.49\linewidth}
        \centering
        \includegraphics[width=\linewidth]{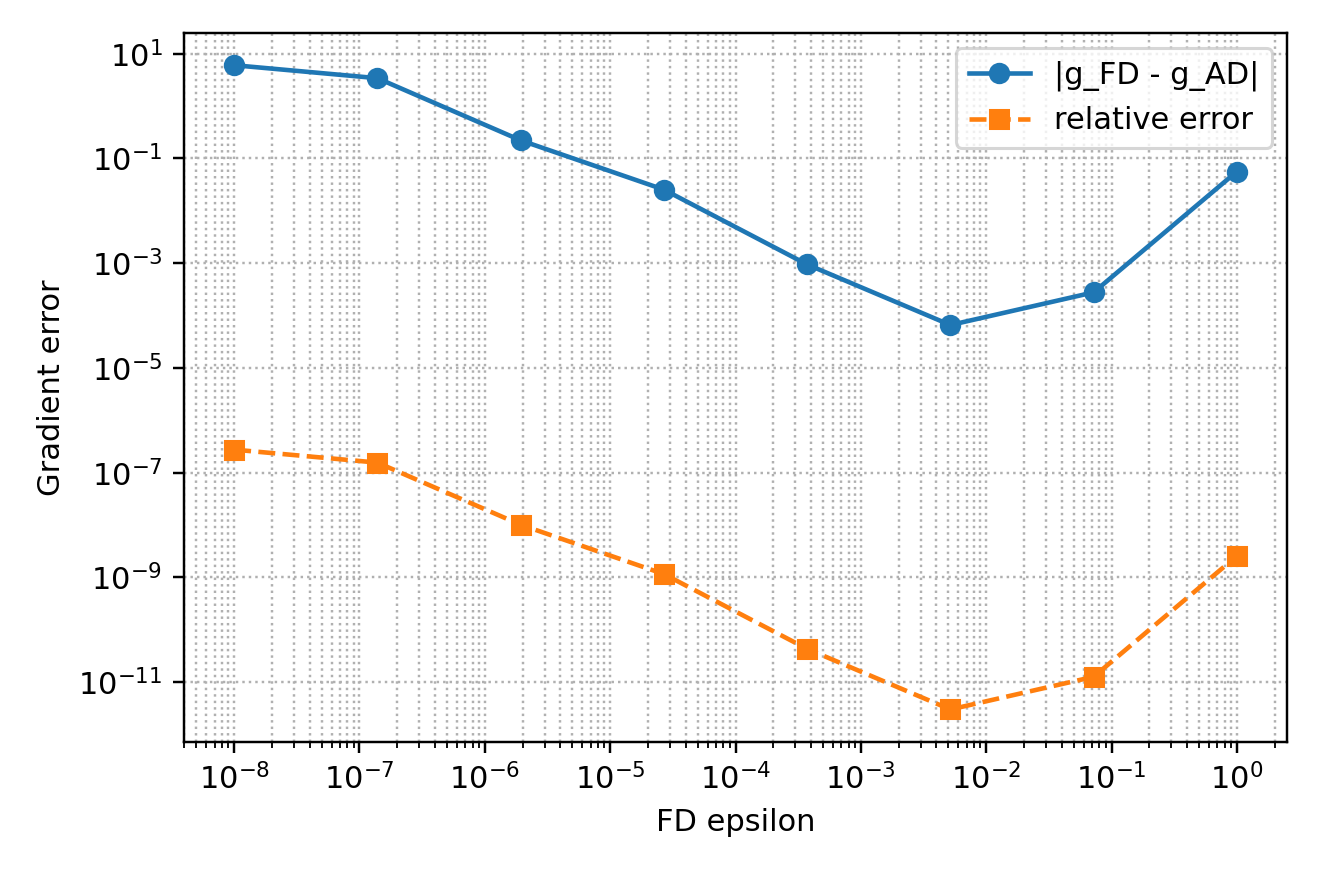}
        \caption{$G_{\mathrm{shear},0}$}
    \end{subfigure}
    \caption{Finite-difference verification of the adjoint gradients for the uniform thermal-preconditioning problem. The curves compare the central finite-difference gradient with the adjoint gradient for the two active inverse controls. The relative-error minima occur at intermediate perturbation sizes, confirming the adjoint sensitivities used in the optimization.}
    \label{fig:fd_syn1}
\end{figure}

Figure~\ref{fig:fd_syn1} shows the finite-difference checks for the two active controls in the uniform thermal-preconditioning problem.

\clearpage

\subsection{Synthetic Problem 2}\label{app:fd_syn2}

\begin{figure}[H]
    \centering
    \begin{subfigure}[b]{0.49\linewidth}
        \centering
        \includegraphics[width=\linewidth]{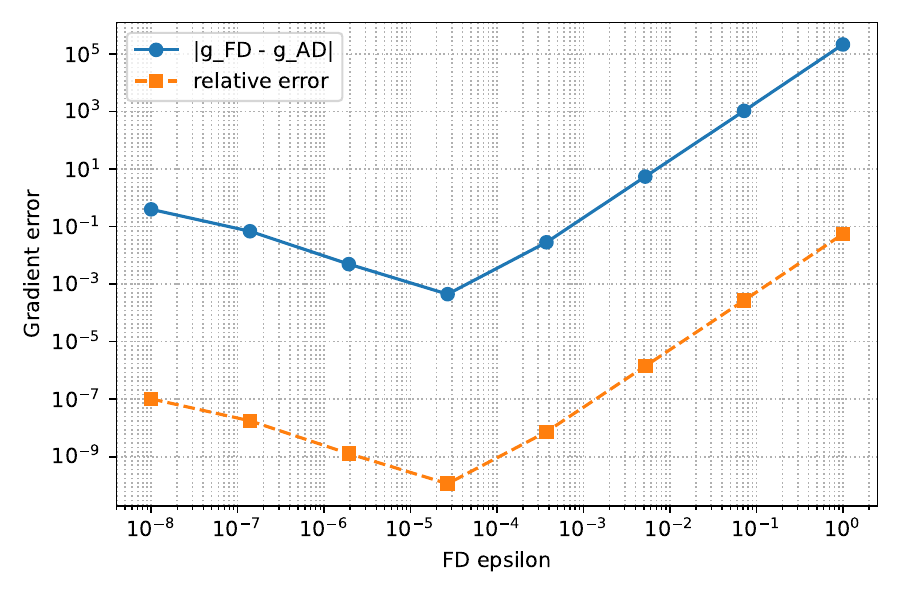}
        \caption{$\alpha$}
    \end{subfigure}\hfill
    \begin{subfigure}[b]{0.49\linewidth}
        \centering
        \includegraphics[width=\linewidth]{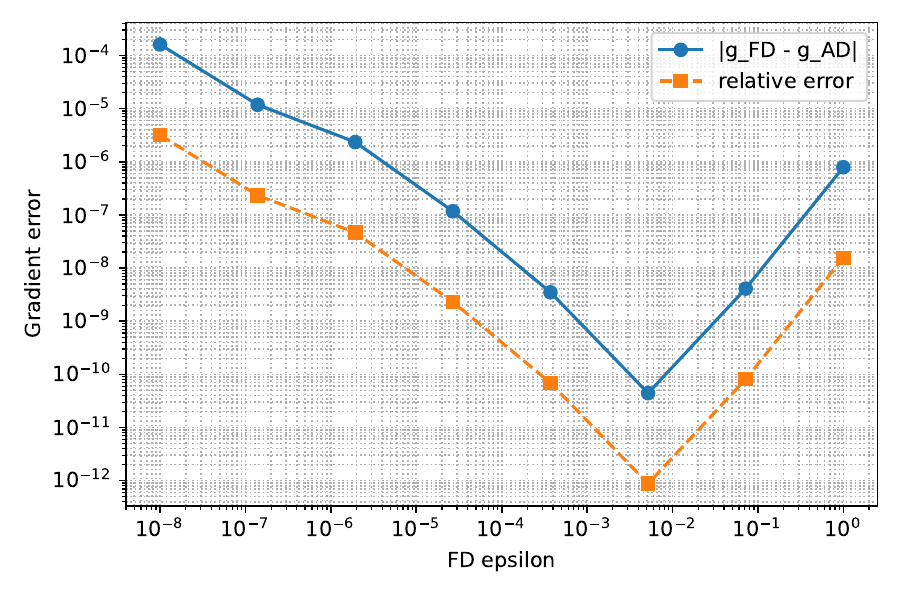}
        \caption{$G_{\mathrm{shear},0}$}
    \end{subfigure}

    \begin{subfigure}[b]{0.49\linewidth}
        \centering
        \includegraphics[width=\linewidth]{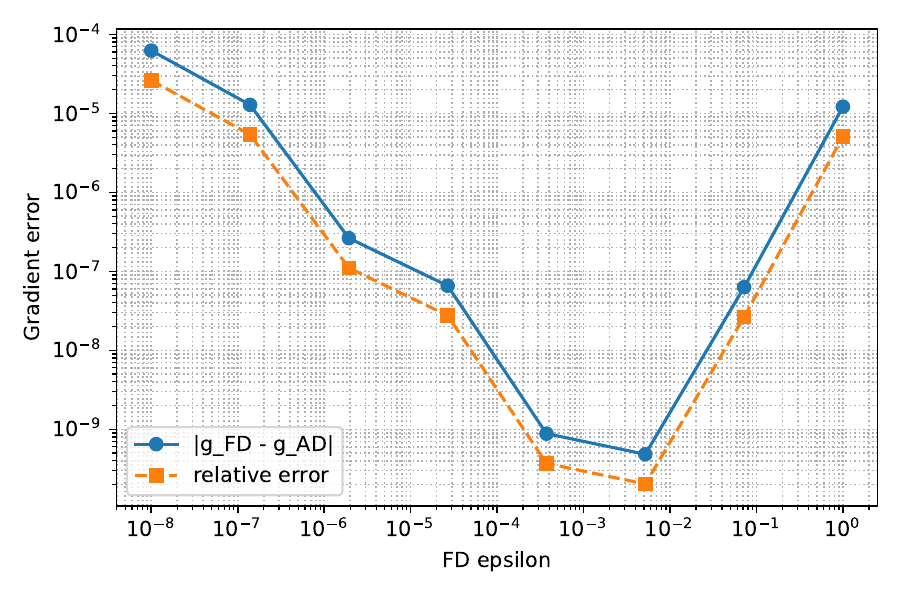}
        \caption{$k_{\mathrm{therm}}$}
    \end{subfigure}

    \caption{Finite-difference verification of the adjoint gradients for the staged rod-contact problem. The checks are performed for the active inverse controls using the same inverse objective and boundary-condition treatment as the reported optimization, with a reduced representative time history for computational efficiency. The finite-difference and adjoint gradients agree over the expected intermediate perturbation range.}
    \label{fig:fd_syn2}
\end{figure}

Figure~\ref{fig:fd_syn2} shows the corresponding finite-difference checks for the active controls in the staged rod-contact problem.

\clearpage

\subsection{Experimental Data}\label{app:fd_exp}

\begin{figure}[H]
    \centering
    \begin{subfigure}[b]{0.49\linewidth}
        \centering
        \includegraphics[width=\linewidth]{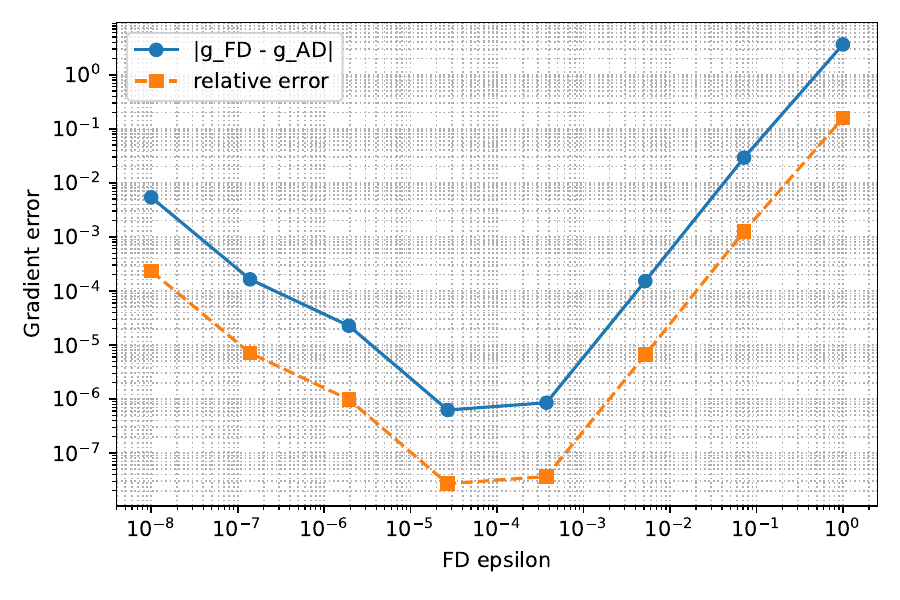}
        \caption{$\alpha$}
    \end{subfigure}\hfill
    \begin{subfigure}[b]{0.49\linewidth}
        \centering
        \includegraphics[width=\linewidth]{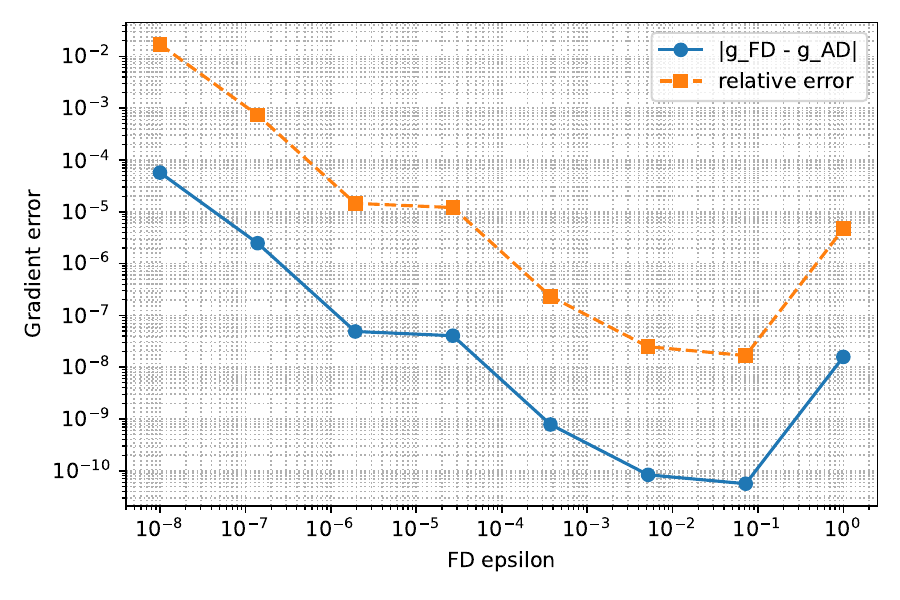}
        \caption{$\beta_x$}
    \end{subfigure}

    \begin{subfigure}[b]{0.49\linewidth}
        \centering
        \includegraphics[width=\linewidth]{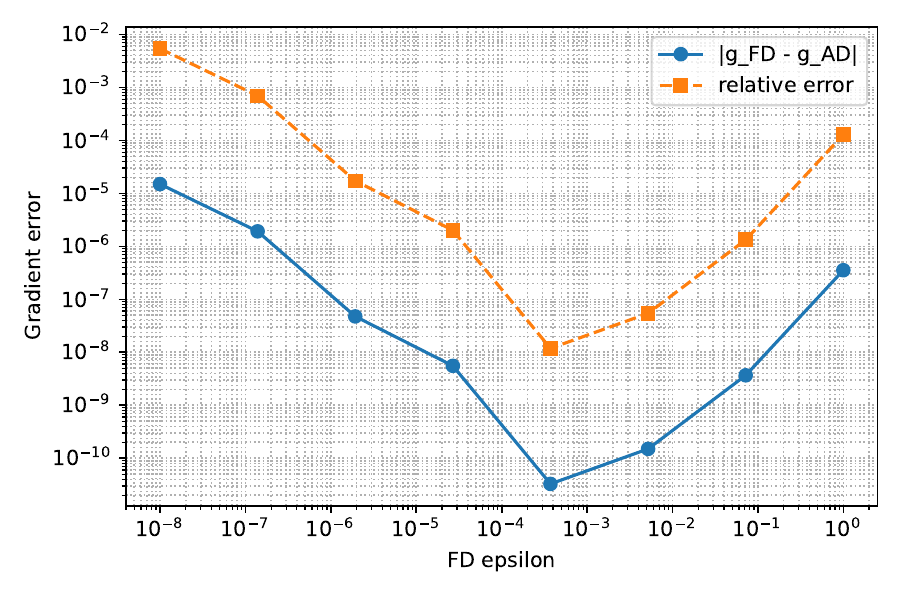}
        \caption{$\beta_y$}
    \end{subfigure}

    \caption{Finite-difference verification of the adjoint gradients for the experimental thermal inverse problem. The checks are performed for the thermal expansion coefficient and the two directional shrinkage amplitudes using the same objective and boundary-condition treatment as the reported experimental optimization. The error curves compare central finite differences from fresh objective evaluations with the adjoint gradients and show agreement over the expected intermediate perturbation range.}
    \label{fig:fd_exp}
\end{figure}

Figure~\ref{fig:fd_exp} shows the finite-difference checks for the three active controls in the experimental thermal inverse problem. The checks use the same reduced experimental objective as the reported optimization, including the Robin-type top-surface temperature assimilation and the two in-plane force-history terms.

\end{document}